\documentclass[useAMS,usenatbib]{mn2e}
\usepackage[english]{babel}
\usepackage{amsmath}
\usepackage{graphicx}
\usepackage{subfigure}

\usepackage[]{natbib}
\title[Modelling AGN in cosmological simulations]{A refined sub-grid model for black hole accretion and AGN feedback in large cosmological simulations}
\author[Steinborn et al.]{Lisa K. Steinborn$^{1}$\thanks{E-mail: steinborn@usm.lmu.de}, Klaus Dolag$^{1,2}$, Michaela Hirschmann$^{3}$, M. Almudena Prieto$^{4,5}$, \and Rhea-Silvia Remus$^{1}$
\\
$^{1}$Universit\"ats-Sternwarte M\"unchen, Scheinerstr.1, D-81679 M\"unchen, Germany\\
$^{2}$Max-Plank-Institut f\"ur Astrophysik, Karl-Schwarzschild Strasse 1, D-85740 Garching, Germany\\
$^{3}$UPMC-CNRS, UMR7095, Institut d'Astrophysique de Paris, Boulevard Arago, F-75014 Paris, France\\
$^{4}$Instituto de Astrof\'{i}sica de Canarias (IAC), V\'{i}a L\'{a}ctea s/n, La Laguna, E-38200, Spain\\
$^{5}$Departamento de Astrof\'{i}sica, Facultad de F\'{i}sica, Universidad de La Laguna, Astrof\'{i}sico Fco. S\'{a}nchez s/n, La Laguna, E-38207, Spain}
\begin{document}

\date{Accepted 2015 January 13. Received 2014 December 16; in original form 2014 September 10}

\pagerange{\pageref{firstpage}--\pageref{lastpage}} \pubyear{2014}

\maketitle

\label{firstpage}

\begin{abstract}
In large scale cosmological hydrodynamic simulations simplified
sub-grid models for gas accretion onto black holes and AGN feedback
are commonly used. 
Such models typically depend on various free parameters, which are not  
well constrained. 
We present a new advanced model containing a more detailed description
of AGN feedback, where those parameters reflect the results of recent
observations.  
The model takes the dependency of these parameters on the black hole
properties into account and describes a continuous transition between
the feedback processes acting in the so-called radio-mode and
quasar-mode.  
In addition, we implement a more detailed description of the accretion
of gas onto black holes by distinguishing between hot and cold gas
accretion. 
Our new implementations prevent black holes from gaining too much
mass, particularly at low redshifts,
so that our simulations are 
successful in reproducing the observed present-day black hole mass function.
Our new model also suppresses star formation in 
massive galaxies slightly more efficiently than many state-of-the-art models.
Therefore, the simulations that include our new implementations
produce a more realistic population of quiescent and star-forming
galaxies compared to recent observations, even if some discrepancies
remain.  
In addition, the baryon conversion efficiencies in our simulation are
-- except for the high mass end --
consistent with observations presented in literature over the mass
range resolved by our simulations.
Finally, we discuss the significant impact of the feedback model
on the low-luminous end of the AGN luminosity function.
\end{abstract}

\begin{keywords}
black hole physics, methods: numerical, galaxies: active, galaxies:
evolution, galaxies: nuclei, quasars: supermassive black holes 
\end{keywords}

\section{Introduction}
Black holes play an essential role in the formation and evolution of galaxies.
They can even influence galaxy clusters and the intra cluster medium (ICM).
However, observations of active galactic nuclei (AGN) indicate that
gas accretion onto black holes and AGN feedback are complex processes,
which are not yet fully understood (e.g. \citealt{Merloni_Heinz},
\citealt{McNamara}, \citealt{Ma}). 
There is evidence for two distinct phases of AGN activity and
feedback: the radio-mode and the quasar-mode. 
The radio-mode is characterized by large radio jets generating hot
X-ray cavities (\citealt{Russell}, \citealt{Mezcua}), whereas in the
quasar-mode the emission is dominated by the accretion disc, which is
visible as the so-called blue bump in the spectrum of quasars and
Seyfert galaxies (e.g. \citealt{Elvis}, \citealt{Prieto}). 

\citet{Churazov} characterized this distinction in a theoretical model
by describing AGN feedback with two components: radiation and
mechanical outflow. 
In their model the amount of energy associated with each component
depends on the Eddington ratio
$f_{\mathrm{Edd}}=\dot{M_\bullet}/\dot{M}_{\mathrm{Edd}}$. 
When a black hole accretes with the Eddington accretion rate
$\dot{M}_{\mathrm{Edd}}$, gas cooling and AGN feedback are in
equilibrium. 
\citet{Churazov} also took advection-dominated accretion
flows (ADAFs) into account, although a jet contribution can
successfully replace an ADAF (\citealt{Falcke},
\citealt{Fernandez_Ontiveros}).   

To constrain this model and to really understand the origin of
different types of AGN and how they influence their environment, large
cosmological simulations play a key role.
They have two major advantages: firstly, they provide a statistically
large sample of black holes. 
This allows to compare the simulations to the newest and
currently most complete observations of the $M_{\bullet}$-$M_*$
relation (e.g. \citealt{McConnell}) or black hole mass functions
(e.g. \citealt{Marconi}, \citealt{Shankar04}, \citealt{Shankar09}) and
stellar mass functions (e.g. \citealt{Muzzin}, \citealt{Bernardi}), in
particular the very massive end. Secondly, having large enough
cosmological boxes where also massive galaxy clusters form, allows to
probe the influence of black holes across all scales of cosmic
environment.  

There already exist a number of studies discussing large cosmological
simulations that include black holes (e.g. \citealt{DiMatteo05},
\citealt{DiMatteo08}, \citealt{Robertson}, \citealt{Teyssier},
\citealt{DeGraf_2010}, \citealt{Booth}, \citealt{Khandai},
\citealt{Rosas_Guevara}, \citealt{Hirschmann},
\citealt{Vogelsberger_nature}, \citealt{Schaye}). 
Those simulations mostly use the black hole model implemented by
\citet{Springel_BHs} or are based on it. In these models -- in contrast
to some more simplified black hole models (e.g. \citealt{Battaglia})
-- black holes are typically described as sink particles which have
fundamental properties like mass and accretion rate, which can be
linked directly to observables. 
Hence, we can study black hole growth and the co-evolution between
black holes and their host galaxies to constrain  and improve the
parametrization of the underlying model. 
In the model from \citet{Springel_BHs} the gas accretion onto black
holes is calculated according to the Bondi formula (\citealt{Hoyle},
\citealt{Bondi}, \citealt{Bondi_Hoyle}), multiplied by a so-called
boost factor $\alpha$. This factor was introduced to account for the
limited resolution in simulations leading to smaller densities and
larger temperatures near the black hole \citep{Booth}. 
To estimate the AGN feedback, a constant value for the radiative
efficiency is typically used (\citealt{Shakura}).

For low resolutions this model works reasonably well.
However, to study not only the origin
of the observed fundamental relations between black holes and their
host galaxies (\citealt{Haering}, \citealt{Tremaine},
\citealt{McConnell}), but also the impact of gas accretion and AGN
feedback on the morphology of the galaxy, simulations with higher
resolution are needed. Until now, this was only studied in simulations
of isolated galaxies and mergers of galaxies (e.g. by
\citealt{Hopkins_merger}, \citealt{DeBuhr}, \citealt{Wassenhove},
\citealt{Capelo}) as well as in cosmological zoom simulations (e.g. by
\citealt{Angles}, \citealt{Marinacci}, \citealt{Dubois_zoom},
\citealt{Choi_2014}). To reproduce both statistical black hole and
galaxy properties within a fully cosmological context and across
various environments in a statistically relevant sample size,
large cosmological boxes with high resolution are needed.
This is still a challenge, but thanks to increasing computational power
it now becomes feasible.
However, despite of this success, new challenges arise as simulations
typically over-estimate the high-mass end of the black hole
and stellar mass function
(e.g. \citealt{Sijacki_2014}, \citealt{Khandai}, \citealt{Vogelsberger14},
\citealt{Genel}, \citealt{Hirschmann}).
Therefore, a more detailed black hole model is necessary.

In this work we extend the model by \citet{Springel_BHs} by improving
the treatment of the two modes of AGN feedback: radiation and
mechanical outflows. 
Following theoretical predictions (\citealt{Churazov},
\citealt{White}, \citealt{Narayan}) as well as recent observational
results (\citealt{Davis}, \citealt{Chelouche}, \citealt{Russell})
gives us estimates for the corresponding two efficiencies depending on
the black hole mass and the accretion rate, which outreaches the
simplified black hole model commonly used in simulations. 

Following \citet{Sijacki}, a steep transition between radio-mode and
quasar-mode is often used in current simulations
(e.g. \citealt{Fabjan}, \citealt{Hirschmann}). 
This is only a rough approximation to the smooth transition which is
observed and also theoretically expected.  
Adopting the model by \citet{Churazov} - which was already constrained
by observations, e.g. \citealt{Russell} - allows us to get a smooth
transition between the two modes. 
This was used by \citet{Hirschmann} to calculate AGN
luminosities, but it was never implemented into simulations. 
Such modifications were also suggested by a recent paper of
\citet{Sijacki_2014}, who studied the AGN luminosity function within a
cosmological simulation using a constant radiative efficiency. They
concluded that in the radio-mode radiative efficiencies might depend
on the accretion rate and on average should be lower than the value
0.1 used in the original black hole model from \citet{Springel_BHs}. 
Furthermore, \citet{Davis} and \citet{Chelouche} found that the
radiative efficiency not only correlates with the accretion rate, but
also with the black hole mass. 

Another deficiency in current implementations of black holes in
cosmological simulations is that the (original) Bondi model predicts
far too low accretion rates during the quasar-mode so that black
holes do not reach the observed masses for a given bulge mass.  
Therefore, a so-called boost factor is commonly used to artificially
raise the accretion rates. This results in realistic accretion rates
for the accretion of cold gas.
However, it has the disadvantage
that it also raises the accretion rate when the hot gas content is
large enough to fulfil the assumptions of the Bondi model, namely when the
gas is distributed in an isotropic sphere. 
This typically is the case in old quiescent galaxies. Consequently,
black holes become too massive at low redshifts. Hence, accretion
rates have to be lower in the radio-mode \citep{Li_Ostriker_Sunyaev}. 

Indeed, several studies adapt the black hole model for higher
resolution simulations by using a boost factor which depends on the
resolution (\citealt{Choi}, \citealt{Choi_2013}), density
(\citealt{Booth}), pressure (\citealt{Vogelsberger}) or angular
momentum \citep{Rosas_Guevara}, although none of them contains a
direct distinction between the accretion of cold and hot gas, 
even if the existence of such two distinct accretion modes has been
shown by observations (e.g. \citealt{Hlavacek-Larrondo}) and predicted
by high-resolution simulations of black hole accretion on sub-kpc
scales (\citealt{Gaspari}, \citealt{Bourne}) as well as
semi-analytical models (e.g. \citealt{Somerville},
\citealt{Hirschmann12}, \citealt{Fanidakis11},
\citealt{Fanidakis_accretionmodes}).  
A distinction between accretion of cold and hot gas
based on the multi-phase model from \citet{Springel_Hernquist}
was implemented in the simulations from \citet{Pelupessy}.
In their study, the molecular gas of the star forming particles was
evaluated from a multi-phase model, in which the accretion of this cold
gas was evaluated separately without any boost factor, assuming the
corresponding temperature as fixed in the underlying multi-phase model.  
 
A black hole mainly grows in the quasar-mode, where cold gas forms an
accretion disc around the black hole which leads to higher accretion
rates. During that period, black holes grow until the AGN feedback and
gas cooling are in equilibrium. At that point, they reach the
$M_{\bullet}$-$\sigma$ relation \citep{Churazov} and thus, the
$M_{\bullet}$-$M_*$ relation. Consequently, the accretion rate
drops until the black hole crosses the threshold towards the
radio-mode. As reviewed by several authors
(e.g. \citealt{Yuan_Narayan}, \citealt{Heckman}), the accretion in the
radio-mode, sometimes also called jet-mode, can be described with
ADAFs containing hot gas \citep{Yuan}. Alternatively, the accretion of
hot adiabatic gas can be described with the Bondi model
\citep{Gaspari}.  Therefore, we distinguish between hot and cold gas
and estimate the accretion rate separately for both gas phases. This
allows us to use different boost factors for hot and cold gas and thus,
to account for both observed accretion modes.  

The outline of this paper is as follows: in section
\ref{Theoretical_model} we describe our black hole model. The set-up
of the cosmological simulations is presented in section
\ref{Simulations}. In section \ref{Results}, adopting different models
for black hole accretion and AGN feedback, we show the results for
our simulations, in particular the evolution of the black hole mass,
the stellar mass and the star formation rate. In section
\ref{discussion} we discuss the radiative efficiency in the radio-mode
and its influence onto the AGN luminosity functions.
Furthermore, we compare our results with other cosmological simulations.
Finally, in section \ref{Summary}, we summarize our main results.  

\section{Theoretical Model}
\label{Theoretical_model}
\subsection{Black hole accretion}
The Bondi model is commonly used in simulations to estimate the black
hole accretion rate. The Bondi accretion rate $\dot{M}_\mathrm{B}$
(\citealt{Bondi}, \citealt{Shima}) is given by \begin{equation}
\dot{M}_\mathrm{B} = \frac{4 \pi G^2 M_\bullet^2 \rho_{\infty}}{(v^2 +
  c_{\mathrm{s}}^2)^{3/2}}, 
\label{accretion_rate}
\end{equation}
where $M_{\bullet}$ is the black hole mass, $\rho$ is the
density, $c_s$ is the sound speed of the accreted gas and $v$ is the
velocity of the gas relative to that of the black hole. Since
\citet{Bondi} assumed an isotropic and isothermal sphere of gas for
his estimation, it is not straight forward to adopt this Bondi accretion
model for hydrodynamic, cosmological simulations aiming to
follow a self consistent accretion history of black holes. For the
implementations based on \citet{Springel_BHs}, the accretion rate of
the black hole is estimated by 
\begin{equation}
\dot{M}_{\mathrm{B}} = \frac{4 \pi \alpha G^2 M_{\bullet}^2
  \left\langle \rho \right\rangle}{(\left\langle c_\mathrm{s}
  \right\rangle^2 + \left\langle v \right\rangle ^2)^{3/2}}, 
\label{accretion_rate_mean}
\end{equation}
where $\left\langle\rho\right\rangle$, $\left\langle v\right\rangle$
and $\left\langle c_s\right\rangle$ are computed using  kernel
weighted SPH estimations. 
Due to limited numerical resolution in such
simulations, the original equation (\ref{accretion_rate}) is
multiplied by a boost factor $\alpha$, which in \citet{Springel_BHs} is
set to a value of $\alpha=100$. Note that the SPH estimates
also depend on the type of SPH kernel and the number of neighbours.
To make this estimation less sensitive to the actual structure
of the multi phase media in the vicinity of the black hole and
therefore the algorithm less dependent on resolution and
on the actual choice of numerical parameters for the kernel weighted
interpolation, \citet{Choi} suggested to use a different way of
building the averages:
\begin{equation}
\dot{M}_{\mathrm{B}} = \left\langle \frac{4 \pi \alpha G^2
    M_{\bullet}^2 \rho}{(c_\mathrm{s}^2 + v^2)^{3/2}}\right\rangle . 
\label{accretion_rate_mean_AA}
\end{equation}
Still, choosing the correct value for the boost factor $\alpha$ is not trivial.
Since due to the limited resolution the density in the not resolved
vicinity of black holes is large, it will be underestimated
and -- in turn -- the temperature (and thus the sound speed) will be
overestimated. Following this argument, \citet{Booth} parametrize
$\alpha$, which is chosen to be $\alpha=1$  as long as the density is
below the critical value where one can assume the gas to be in the
hot phase. For larger densities, when gas is accreted mainly in a cold
phase, $\alpha$ increases with density. Alternatively,
\citet{Vogelsberger} have presented a recipe for modelling $\alpha$ based
on the equilibrium between cooling losses and AGN feedback. However,
both models do not directly account for the different accretion modes
of hot and cold gas phase, where cold gas usually is accreted in
turbulent streams, whereas hot gas indeed can be assumed to be
isotropic and isothermal. 

In our model, we use a sixth-order Wendland kernel
\citep{Dehnen} with 295 neighbours, building the mean values according
to equation (\ref{accretion_rate_mean}) and directly distinguishing
between the accretion of hot and cold gas.
In this way, we can safely use the original estimate of building
the averages, which has the advantage to be more sensitive to density
structures close to the black hole. In general, we assume hot gas has
temperatures above $T \approx 10^6 K$, whereas cold gas has
temperatures below $T \approx 10^5 K$ \citep{Gaspari}. Since we do not
account for a third warm phase, we choose $T=5 \cdot 10^5$K as
threshold between hot and cold gas.
In contrast to \citet{Pelupessy}, who use the molecular
fraction of the gas for star-forming particles from the multi-phase
model \citep{Springel_Hernquist} to account for cold gas accretion,
we also assign gas with a temperature below our threshold in
addition to the star forming gas to the cold phase. 
For both gas phases the accretion
rate is calculated separately according to equation
\ref{accretion_rate_mean}, but with different values for $\alpha$ 
according to the result by \citet{Gaspari}, who argue that due to
turbulence the assumptions of the Bondi model are not fulfilled for
the cold gas.
When they include cooling and turbulence in their simulation, they find
an accretion rate which is around 100 times larger than the Bondi
accretion rate. 
Interestingly, this is the same value which is used as boost factor
$\alpha$ in the original model from \citet{Springel_BHs}. 
But for adiabatic accretion, the difference, \citet{Gaspari} find, is
about one order of magnitude smaller.  
Hence, we use $\alpha=10$ for hot gas and $\alpha=100$ for cold gas.  

Furthermore, the black hole accretion rate $\dot{M_\bullet}$ is
limited to the Eddington accretion rate 
\begin{equation}
\label{Mdot_Edd}
\dot{M}_{\mathrm{Edd}} = \frac{4 \pi G M_{\bullet}
  m_\mathrm{p}}{\eta_{\mathrm{Edd}} \sigma_\mathrm{T} c}, 
\end{equation}
where $m_\mathrm{p}$ is the proton mass, $\sigma_\mathrm{T}$ the
Thompson scattering cross section and $\eta_{\mathrm{Edd}}$ the
feedback efficiency if the black hole would accrete with
$\dot{M}_{\mathrm{Edd}}$. 
Then the accretion rate is given by:
\begin{equation}
\dot{M_\bullet} = \mathrm{min}(\dot{M}_\mathrm{B, hot} +
\dot{M}_\mathrm{B, cold}, \dot{M}_{\mathrm{Edd}}) . 
\end{equation}
The distinction between hot and cold gas accretion leads to a faster
black hole growth in the quasar-mode, because when
calculating the mean value of the sound speed $\left\langle c_s
\right\rangle$ and the gas velocity $\left\langle v \right\rangle$
only for cold gas, the accretion rate estimated with equation
(\ref{accretion_rate_mean}) is higher than calculating the mean values
of both cold and hot gas together. This solves the well known problem
of too low gas accretion, which was addressed in other
simulations by increasing the maximum accretion rate to a few times
$\dot{M}_{\mathrm{Edd}}$ (e.g. \citealt{DiMatteo12}), which is not
needed in our simulations. 

\subsection{AGN feedback}
In the commonly used black hole model by \citet{Springel_BHs}, the
feedback energy per unit time is calculated as 
\begin{equation}
\dot{E} = \epsilon_\mathrm{f} \epsilon_\mathrm{r} \dot{M_\bullet} c^2,
\label{feedback_energy_old}
\end{equation}
where $\epsilon_\mathrm{f}$ is the efficiency with which the energy
radiated from the black hole is coupled to the ISM
(\citealt{Springel_BHs}, \citealt{Booth}) and $\epsilon_\mathrm{r}$ is
the radiative efficiency.  

The original model as used in \citet{Hirschmann} is simplified, since
it uses a constant radiative efficiency and thus does not allow for a
smooth transition between quasar- and radio-mode. Furthermore, it
neglects mechanical feedback, which was already implemented in other
simulations as AGN driven winds (i.e. \citealt{Choi_2014}). 
To account for both mechanical and radiative feedback, we adopt a new
feedback scheme based on \citet{Churazov}. In this study, they propose
that AGN feedback can be split up into two components: 
\vspace{0.2cm}
\begin{enumerate}
\item{\textbf{Outflow:} The outflow component is a mechanical
    feedback which dominates at accretion rates below $\sim
    0.01\dot{M}_{\mathrm{Edd}}$ and diminishes at accretion rates above $\sim
    0.1\dot{M}_{\mathrm{Edd}}$. The corresponding gas heating power is given
    by:} 
\begin{equation}
P_\mathrm{o} = \epsilon_\mathrm{o} \dot{M_\bullet} c^2,
\label{outflow_power}
\end{equation}
where $\epsilon_\mathrm{o}$ is the outflow efficiency.
\vspace{0.2cm}
\item{\textbf{Radiation:} The radiative component dominates near the
    Eddington limit ($f_{\textrm{Edd}}>0.1$) and has the luminosity} 
\begin{equation}
L=\epsilon_\mathrm{r} \dot{M_\bullet} c^2.
\label{radiation_power}
\end{equation}
\end{enumerate}
\vspace{0.2cm}
\begin{figure}
  \includegraphics[trim = 10mm 0mm 22mm 10mm, clip,
  width=0.5\textwidth]{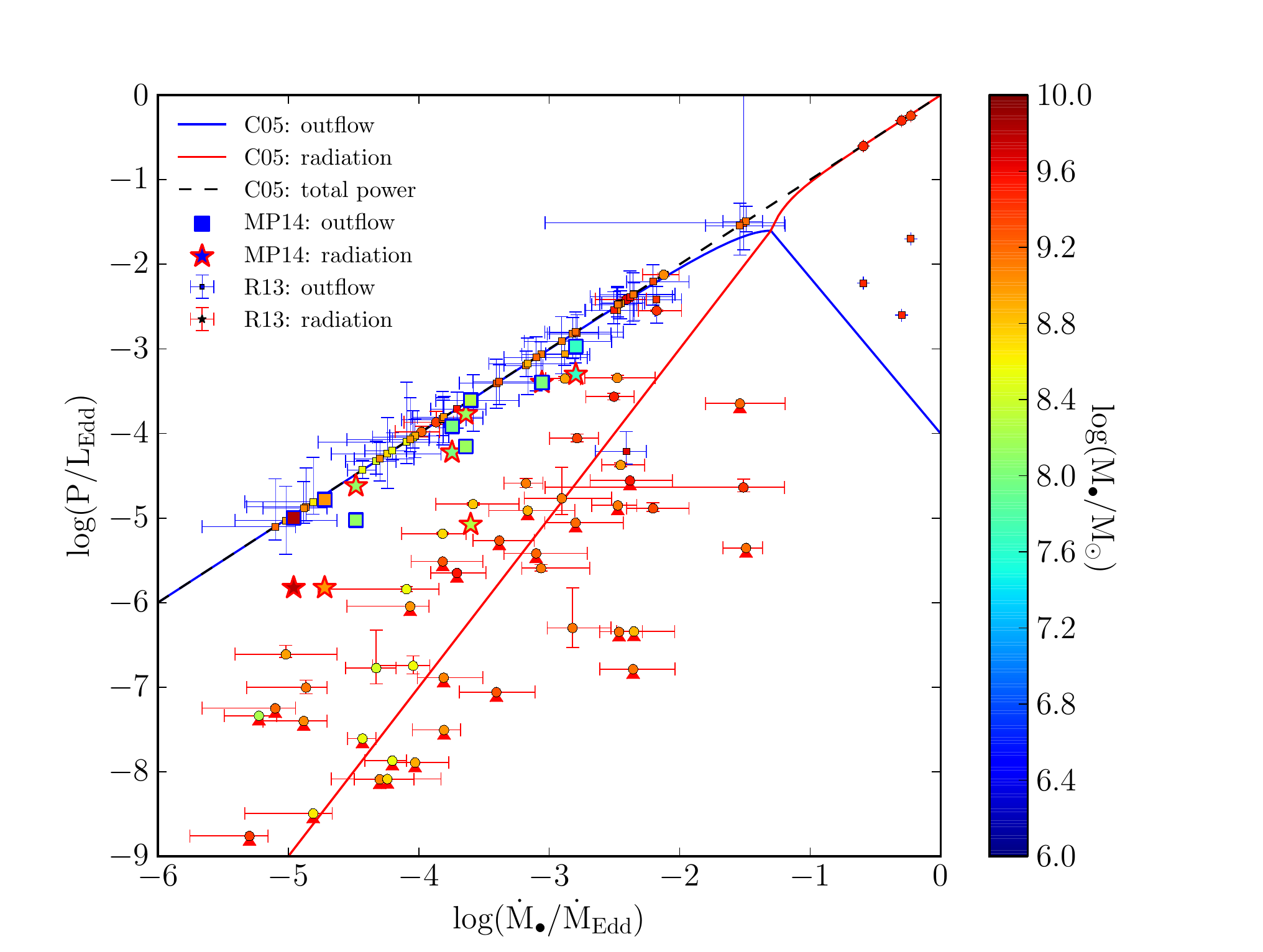} 
  \caption{The lines show the predictions by \citealt{Churazov}
    (C05) for the power of the radiation (red line), the mechanical
    outflow (blue line) and the sum of both (black dashed
    line). Observations of jet powers (blue errorbars and
      edges) and luminosities (red errorbars and edges)
    constrain the difference between both components. This figure
    includes two different observations: The big stars and squares
    show recent observations by \citealt{Mezcua} (MP14) and the data
    with blue and black errorbars are observations by \citealt{Russell}
    (R13). Black triangles mark upper limits. Furthermore, the black
    hole masses are indicated by the colors of the symbols. Since the
    masses used by R13 are based on K-band magnitudes, which are known
    to be inaccurate, we used the dynamical masses by \citet{McConnell}
    for the sources included in both samples.}
\label{Churazov_model_plot}
\end{figure}

We implement both radiative and mechanical AGN feedback as thermal
feedback due to the inability to resolve the sub-kpc scales, where the
jets provide the mechanical feedback.  
The feedback energy per unit time in this model is then the sum of
$P_{\mathrm{o}}$ and the fraction $\epsilon_\mathrm{f}$ of the
luminosity: 
\begin{equation}
\dot{E} = (\epsilon_\mathrm{o} + \epsilon_\mathrm{f}
\epsilon_\mathrm{r}) \dot{M_\bullet} c^2. 
\label{feedback_energy_new}
\end{equation} 
The effect of accreted matter can be split into outflow and radiation
components: 
\begin{equation}
\frac{\dot{M_\bullet}}{\dot{M}_{\mathrm{Edd}}} =
\frac{P_\mathrm{o}}{L_{\mathrm{Edd}}} + \frac{L}{L_{\mathrm{Edd}}}, 
\label{mdot_splitted}
\end{equation}
where the Eddington accretion rate
\begin{equation}
\dot{M}_{\mathrm{Edd}} = \frac{L_{\mathrm{Edd}}}{\eta_{\mathrm{Edd}} c^2}
\label{Mdot_Edd_Toy}
\end{equation}
depends on the total efficiency
\begin{equation}
\eta := \epsilon_\mathrm{o} + \epsilon_\mathrm{r}.
\label{eta_sum}
\end{equation}
This model is shown as solid lines (blue corresponds to mechanical
outflow and red to radiation) in Fig. \ref{Churazov_model_plot}, 
which were adopted from \citet{Churazov}. For the outflow-dominated
regime they assume 
\begin{equation}
\frac{L}{L_{\mathrm{Edd}}} = 10 \cdot \left(\frac{\dot{M_\bullet}}{\dot{M}_{\mathrm{Edd}}}\right)^2
\label{radiation}
\end{equation}
as a lower limit for the radiation, which is a consequence of
advection-dominated accretion flows \citep{Narayan}. 
In the radiation-dominated regime the outflow decreases with the
Eddington ratio: 
\begin{equation}
\frac{P_\mathrm{o}}{L_{\mathrm{Edd}}} = 10^{-4} \cdot
\left(\frac{\dot{M_\bullet}}{\dot{M}_{\mathrm{Edd}}}\right)^{-1.8431}. 
\label{outflow}
\end{equation}
This guarantees that the minimum value for the outflow efficiency is
$\epsilon_\mathrm{o} = 10^{-5}$, which was calculated by
\citet{Churazov} assuming that gas cooling and AGN feedback balance
each other at the Eddington limit. 
We choose $\frac{\dot{M_\bullet}}{\dot{M}_{\mathrm{Edd}}} = 0.05$ as the
threshold between radio and quasar mode. The value for the outflow at
$\frac{\dot{M_\bullet}}{\dot{M}_{\mathrm{Edd}}} = 1$ follows the
calculations of \citet{Churazov}, who find $\epsilon_\mathrm{o}
\approx 10^{-5}$ for black holes accreting with the Eddington
accretion rate. 

The feedback model of \citet{Churazov} was recently confirmed by
observations 
(see also \citealt{Russell}) measuring luminosities and cavity powers
of a large sample of unresolved nuclear X-ray sources. Most of the
selected brightest cluster galaxies (BCGs) have large X-ray
cavities. The data from \citet{Russell} show a large scattering of the
luminosities in 
the radio regime illustrated by round filled circles with black
errorbars in Fig. \ref{Churazov_model_plot}, implying that a secondary
quantity influences the luminosity. A few data points are below the
theoretical lower limit, albeit the uncertainties in the
observations are relatively high. Uncertainties can occur, for
example, when measuring the cavity volume due to projection effects.  
In Fig. \ref{Churazov_model_plot}, the black hole masses are
color-coded as indicated by the colorbar. The masses from
\citet{Russell} are based on K-band magnitudes, which is known to be
problematic. Therefore, we use the dynamical masses from McConnell \&
Ma (2013) for the sources included in both samples. Nearly all
black holes that lie below the prediction are very massive ($>10^9
M_{\odot}$). For lower masses, the observations are in better
agreement with the predictions. We will discuss the uncertainties in
section \ref{The_unknown} in more detail. 

Recently, \citet{Mezcua} presented measurements of luminosities of a
much smaller sample of AGN, but with sufficiently larger angular resolution
and sensitivity. Their estimations for $L_{\mathrm{bol}}$ are more reliable than
those presented in \citet{Russell}, because they measure
$L_{\mathrm{bol}}$ after integrating 
the radio to X-ray Spectral Energy Distribution (SED). Furthermore
they explicitly provide values for 
X-ray cavity powers. For CenA, M87 and NGC1052, they used X-ray
cavities of maser emission from the literature (\citealt{Prieto},
\citealt{Russell}, \citealt{Fernandez_Ontiveros12}). All other values
were estimated using the correlation between core radio luminosity at
5 GHz and $P_{\mathrm{o}}$ of \citet{Merloni_Heinz}. The data from \citet{Mezcua}
is also included in Fig. \ref{Churazov_model_plot}, where the filled
stars represent the luminosities and the squares the cavity powers. 
Since equation (\ref{radiation}) is a lower limit, their luminosities are in
very good agreement with the predictions.
The cavity powers do not always match the blue line,
but as described by \citet{Mezcua}, they are expected to be lower limits, 
because the estimations of $P_o$ do not take into account the energy
which is used to compress the gas when the jet advances the ISM/ICM. 

In simulations, the theoretical and observational results shown in
Fig. \ref{Churazov_model_plot} can be used to calculate the
efficiencies $\epsilon_\mathrm{o}$ and $\epsilon_\mathrm{r}$. 
To estimate the radiative and outflow efficiencies, we first have to
assume a value for the total efficiency $\eta$
and then use the predictions from \citet{Churazov} to separate the
AGN feedback into radiation and mechanical outflow. 
In theoretical studies, the total efficiency is often assumed to be
$0.1$ (e.g. \citealt{Churazov}), however, observations of \citet{Davis}
and \citet{Chelouche} suggest a mass dependence of this parameter. 
In the model from \citet{Churazov}, both $\epsilon_\mathrm{o}$ and
$\epsilon_\mathrm{r}$ 
depend on the accretion rate and the total efficiency. 
For $\dot{M_\bullet}/\dot{M}_{\mathrm{Edd}} < 0.05$ the lower limit
for $\epsilon_\mathrm{r}$ can be calculated with equation
(\ref{radiation_power}) and (\ref{radiation}), i.e. 
\begin{equation}
\epsilon_\mathrm{r, min} = 10 \eta
\frac{\dot{M_\bullet}}{\dot{M}_{\mathrm{Edd}}} 
\label{epsilon_r_radio_min}
\end{equation}
Since this is only a lower limit, all solutions between
$\epsilon_\mathrm{r, min}$ and $\epsilon_\mathrm{r, max} = \eta$
are possible.
Therefore, we introduce the slope $\beta$, which is in
the range between 0 and 1, to get a general expression for
$\epsilon_\mathrm{r}$: 
\begin{equation}
\epsilon_\mathrm{r} = A \cdot \eta
\left(\frac{\dot{M_\bullet}}{\dot{M}_{\mathrm{Edd}}}\right)^{\beta}, 
\label{epsilon_r_radio}
\end{equation}
where $A = 10^{-4} \cdot 0.05^{-2.8431-\beta}$. 
The outflow efficiency is calculated with equation
(\ref{epsilon_r_radio}) and (\ref{eta_sum}). 

\begin{figure}
  \includegraphics[trim = 10mm 0mm 22mm 10mm, clip,
  width=0.5\textwidth]{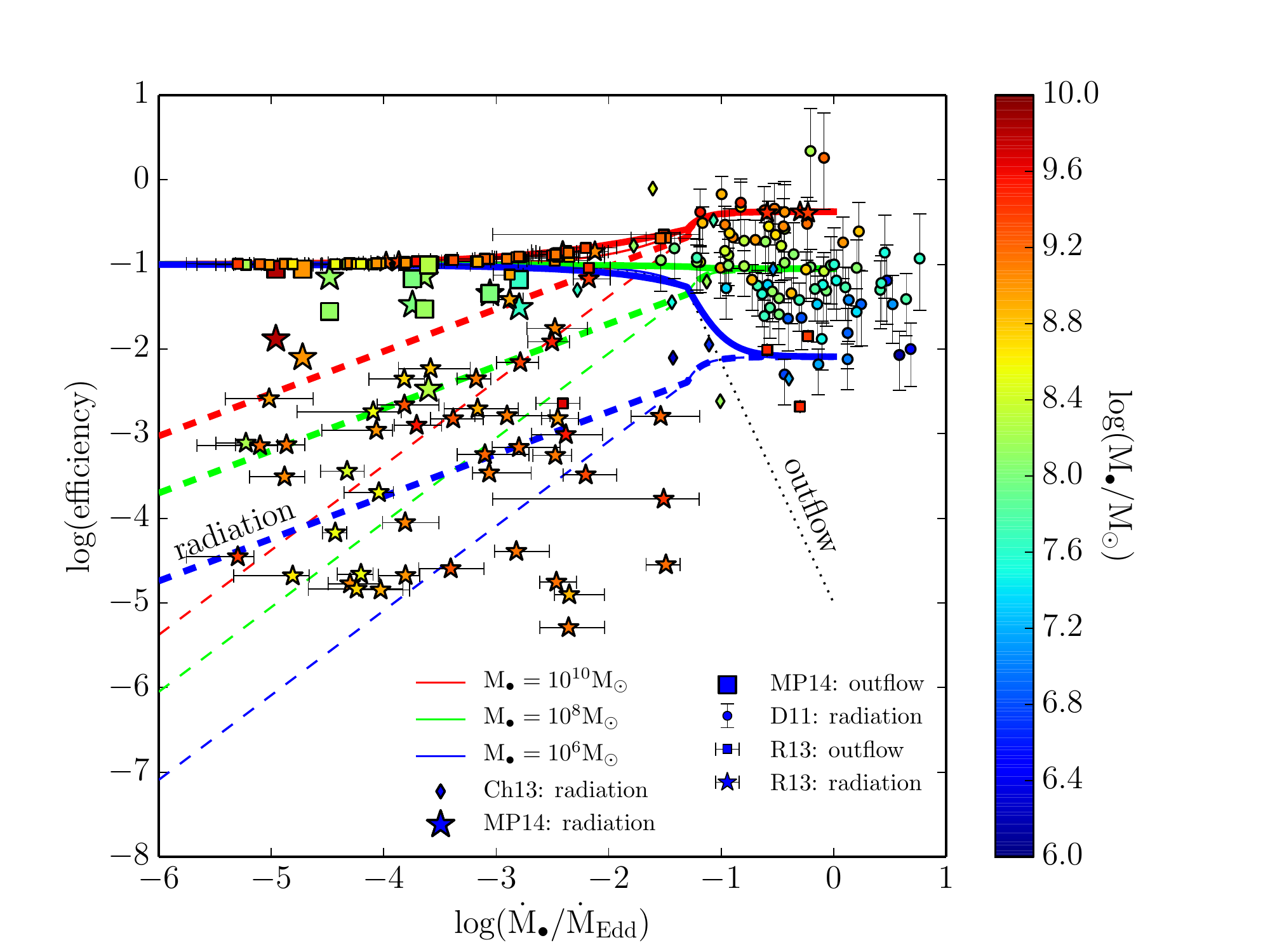}           
  \caption{Our new feedback model includes both outflow (dotted line)
    and radiation (dashed lines) as described by \citet{Churazov}
    as well as a mass dependent radiative efficiency
    following \citet{Davis}. The solid lines show the sum of
    $\epsilon_\mathrm{o}$ and $\epsilon_\mathrm{r}$. The small dots
    and diamonds are observations by \citealt{Davis} (D11) and
    \citealt{Chelouche} (Ch13), who both estimated radiative
    efficiencies. In the radio regime we assume $\eta=0.1$. The large
    stars and squares correspond to recent observations by \citealt{Mezcua}
    (MP14) of the outflow and radiation. From left to
    right the observed galaxies are M87, NGC 4594, NGC 1097, NGC 3169,
    NGC 1386, NGC 2911, NGC 1052 and Cen A. Small stars and squares
    correspond to observations by \citealt{Russell} (R13). The black
    hole masses are color-coded as indicated by the colorbar.
    } 
\label{feedback_model_plot}
\end{figure}

For $\dot{M}/\dot{M}_{\mathrm{Edd}} > 0.05$ the radiation dominates.
The origin of the blue line in Fig. \ref{Churazov_model_plot} in this
regime is the analytical calculation by \citet{Churazov}, which is
based on the equilibrium between gas cooling and heating of gas due to
AGN feedback. Hence, it is not only a lower limit and it is not
necessary to introduce a slope as in the radio regime. In that respect
from equation (\ref{outflow_power}) and (\ref{outflow}) follows 
\begin{equation}
\epsilon_\mathrm{o} = 10^{-4} \eta
\left(\frac{\dot{M_\bullet}}{\dot{M}_{\mathrm{Edd}}}\right)^{-2.8431} 
\label{epsilon_o_quasar}
\end{equation}
and thus $\epsilon_\mathrm{r}=\eta - \epsilon_\mathrm{o}$.
This is shown in Fig. \ref{feedback_model_plot} for different black
hole masses. The filled circles and diamonds in Fig. \ref{feedback_model_plot} are the
observations from \citet{Davis} and \citet{Chelouche} illustrating that
they are not consistent with the model for $\eta=0.1$ (green lines). 
Therefore, we account for the observed spin of black holes by
following the observations of \citet{Davis} for quasars and of
\citet{Chelouche} for Seyfert 1 AGN, who both find a correlation
between the radiative efficiency and the black hole mass. 
Hence, we use the relation found by
\citet{Davis} to estimate the total efficiency at the Eddington limit,
which is approximately the same as the radiative efficiency at the
Eddington limit: 
\begin{equation}
\eta_{\mathrm{Edd}}(M_{\bullet}) \approx \epsilon_\mathrm{r,
  Edd}(M_{\bullet}) = 0.089 \left(\frac{M_{\bullet}}{10^8
    M_{\odot}}\right)^{0.52}. 
\label{eta_var_relation_obs}
\end{equation}
We limit $\eta_{\mathrm{Edd}}(M_{\bullet})$ by the value 0.42, which
is the theoretical maximum efficiency of a rotating black hole.
To calculate the outflow efficiency, the
constant value of $\eta = 0.1$ is used as it is currently difficult to
estimate outflow efficiencies with observations (see section
\ref{The_unknown} for further discussion). 
Equation (\ref{eta_sum}), (\ref{epsilon_r_radio}) and
(\ref{epsilon_o_quasar}) then lead 
to the following set of equations: 
\begin{eqnarray}
\epsilon_{\mathrm{r}} =\begin{cases}
  A \eta_{\mathrm{Edd}}(M_{\bullet})
  \left(\frac{\dot{M_\bullet}}{\dot{M}_{\mathrm{Edd}}}\right)^{\beta},
  \text{ if }\frac{\dot{M_\bullet}}{\dot{M}_{\mathrm{Edd}}} < 0.05, 
\vspace{3mm}\\
  \eta_{\mathrm{Edd}}(M_{\bullet}) - 10^{-4}
  \eta_{\mathrm{Edd}}(M_{\bullet})
  \left(\frac{\dot{M_\bullet}}{\dot{M}_{\mathrm{Edd}}}\right)^{-2.8431},
  \\\text{ otherwise} 
\end{cases}
\label{epsilon_r_final}
\end{eqnarray}
and
\begin{eqnarray}
\epsilon_{\mathrm{o}} =\begin{cases}
  0.1 - A \cdot
  0.1\left(\frac{\dot{M_\bullet}}{\dot{M}_{\mathrm{Edd}}}\right)^{\beta},
  \text{ if }\frac{\dot{M_\bullet}}{\dot{M}_{\mathrm{Edd}}} < 0.05, 
\vspace{3mm}\\
  10^{-5}
  \left(\frac{\dot{M_\bullet}}{\dot{M}_{\mathrm{Edd}}}\right)^{-2.8431},
  \text{ otherwise.} 
\end{cases}
\label{epsilon_o_final}
\end{eqnarray}
In our simulations both radiative and mechanical feedback are
implemented as thermal feedback, 
since we do not resolve jets.

The three coloured lines in Fig. \ref{feedback_model_plot} show the
model from \citet{Churazov} for $\beta=0.5$ (thick dashed lines) and
$\beta=1$ (thin dashed lines) and different black hole masses. The red
lines correspond to $M_{\bullet} = 10^{10} M_{\odot}$, the green ones
to $M_{\bullet} = 10^{8} M_{\odot}$ and the blue ones to $M_{\bullet}
= 10^{6} M_{\odot}$. This is in much better agreement with the
observations than choosing a constant total efficiency. 
In the radio regime, we included observations by \citet{Russell} and \citet{Mezcua},
who measured the power of the radiation and outflow as well as $L_{\mathrm{Edd}}$.
With equation (\ref{mdot_splitted}) they calculated $\dot{M}/\dot{M}_{\mathrm{Edd}}$.
Using the equations (\ref{outflow_power}), (\ref{radiation_power}) and
(\ref{Mdot_Edd_Toy}) we can derive the efficiencies 
\begin{equation}
\epsilon_\mathrm{o} = \eta \cdot \frac{P_0 /
  L_{\mathrm{Edd}}}{\dot{M_\bullet}/\dot{M}_{\mathrm{Edd}}} 
\label{epsilon_o_obs}
\end{equation}
and
\begin{equation}
\epsilon_\mathrm{r} = \eta \cdot \frac{L /
  L_{\mathrm{Edd}}}{\dot{M_\bullet}/\dot{M}_{\mathrm{Edd}}}. 
\label{epsilon_r_obs}
\end{equation}
In the radio regime, it is justified to use $\eta=0.1$. As can be seen
in Fig. \ref{feedback_model_plot}, the data points for the radiative
efficiency do not show the simple trend as assumed in
\citet{Churazov}. In fact, they seem to be consistent with randomly
scattering between $10^{-1}$ and $10^{-5}$. There also seems to be no
mass dependency in the radio regime\footnote{For the data from Russell
  et al. (2013) the dynamical masses from McConnell \& Ma (2013) were
  taken if available. If not, the same masses were taken which Russell
  et al. (2013) used to calculate $L_{\mathrm{Edd}}$.}. 

For NGC 1097 and NGC 1386, the radiation dominates.
The observations by \citet{Mezcua} show that these sources have small
jets, whereas the other sources have larger jets. Interestingly both
NGC 1097 and NGC 1386 have a bar at large scales, but they show no
evidence of a bar on small scales. They both also have a ring of
star-forming regions. This indicates that the morphology of the
galaxies will play a key role for future studies. For simulations this
implies that the resolution has to be high enough to resolve the
morphology of galaxies. Note that this is not the case for the
simulations performed in this work, but will be the aim for
forthcoming studies. 
 
\section{The Simulations}
\label{Simulations}
The present work is based on a set of cosmological simulations called
the Magneticum Pathfinder Simulations\footnote{www.magneticum.org}
(Dolag et al. in prep.). The simulations are performed with an updated
version of the TreePM-SPH code P-GADGET3 \citep{Springel}. 

We adopt a $\Lambda$CDM-cosmology with parameters according to the
seven year results of the Wilkinson Microwave Anisotropy Probe with
$\Omega_m=0.272$, $\Omega_{\Lambda}=0.728$,
$\Omega_{\mathrm{b}}=0.0456$ and $h=0.704$ \citep{Komatsu}. 
We follow the hydrodynamics of the gas using the smoothed particle
hydrodynamics method (see \citealt{Price} for a recent review on the
SPH method). We use an entropy conserving formulation
\citep{Springel_Hernquist02}, where star formation is based on a
multi-phase sub-resolution model by \citet{Springel_Hernquist}. 
Additionally, we include complex treatment for a wide range of
physical processes such as isotropic thermal conduction
\citep{Dolag04} with an efficiency of $\kappa=1/20$ of the classical
{\it Spitzer} value, stellar evolution, metal enrichment and supernova
feedback (\citealt{Tornatore03}, \citealt{Tornatore}), a cooling
function which depends on the individual metal species following
\citet{Wiersma} as well as the treatment of black holes and their
associated feedback based on the model implemented by
\citet{Springel_BHs}. We improve the accuracy, stability and
reliability of our hydrodynamical method with several state-of-the-art
improvements of the SPH method. This includes the higher-order
Wendland kernel functions \citep{Dehnen} as well as time dependent
artificial viscosity to properly track turbulence within galaxy
clusters (\citealt{Dolag05}, \citealt{Donnert}).

Regarding the black hole physics
we use the modifications as described by \citet{Fabjan},
in contrast to the original model implemented by \citet{Springel_BHs},
and made changes to the seeding and further treatment of black holes as
described in detail by \citet{Hirschmann}.
The most important one of these changes is that we do not pin the
black holes to the most bound particles anymore. 
This `pinning' is used in other simulations to keep the black holes in
the centre of their host galaxy, but it also has the side effect that
black holes `jump' from the less massive galaxy to the more massive
one during merger events.
To avoid that the black hole particles are wandering away from the
centre of galaxies by numerical effects, we firstly implemented the
conservation of momentum and centre of mass when two black hole
particles are merging. 
Secondly, we enforce momentum conservation for the smooth accretion of
gas and therefore do not model any momentum transfer when swallowing
gas. 
Without pinning, we have black holes not only in
central galaxies, but also keep them in satellite systems until they
fully merge. Thus, we are able to track black hole growth much better,
in particular in massive galaxy clusters (following all the black holes
in satellite galaxies).

\citet{Hirschmann} already presented a detailed analysis of black hole
growth in the Magneticum Pathfinder Simulations particularly focusing
on the origin of the anti-hierarchical growth of black holes
within a hierarchical structure formation scenario. 
Various observational trends can be already explained using the
simplified black hole model described by \citet{Springel_BHs}. 
However, implementing the more detailed description of AGN feedback
and black hole accretion as described in section
\ref{Theoretical_model} leads to further improvements in predicting
a more realistic population of black holes and AGN in our hydrodynamic
simulations.  
\begin{table*}
\centering
\begin{tabular} {|l|l|l|l|l|l|}
\hline
{} & Box size & initial particle number & $\epsilon_\mathrm{f}$ &
$\epsilon_\mathrm{r}$ & $\epsilon_\mathrm{o}$\\  
{} & [(Mpc/h)$^3$] & {} & {} & {} & {}\\ 
\hline
68Mpc/hr fiducial model & 48$^3$ & $2\cdot 216^3$ & 0.15 & 0.2 & --\\
68Mpc/hr NFM & 48$^3$ & $2\cdot 216^3$ & 0.2 & variable & variable\\
68Mpc/hr NAM & 48$^3$ & $2\cdot 216^3$ & 0.15 & 0.2 & --\\
68Mpc/hr NFAM & 48$^3$ & $2\cdot 216^3$ & 0.2 & variable & variable\\
182Mpc/hr fiducial model & 128$^3$ & $2\cdot 576^3$ & 0.15 & 0.2 & --\\
182Mpc/hr NFAM & 128$^3$ & $2\cdot 576^3$ & 0.2 & variable & variable\\
\hline
\end{tabular}
\caption{General settings of the simulations performed in this
  study. Variable values of $\epsilon_\mathrm{r}$ and
  $\epsilon_\mathrm{o}$ are calculated with equations
  (\ref{epsilon_r_final}) and (\ref{epsilon_o_final}).} 
\label{boxes_resolutions}
\end{table*}

We performed six simulation runs with the same
resolution as in the large $(500 \mathrm{Mpc})^3$ box with an initial
particle number of $2\cdot 1564^3$ analysed 
by \citet{Hirschmann}. In the context of the set of Magneticum
Pathfinder Simulations from Dolag et al. (in prep.) we refer to this
resolution as hr (`high resolution'). 
The particle masses are $M_{\textrm{dm}}=6.9 \cdot 10^8 M_{\odot}/h$,
$M_{\textrm{gas}}=1.4 \cdot 10^8 M_{\odot}/h$ and
$M_{\textrm{stars}}=3.5 \cdot 10^7 M_{\odot}/h$ and the softening
length is 3.75 kpc/h for dark matter and gas and 2.0 kpc/h for stars. 
Black holes are represented as collisionless sink
particles. They are seeded in galaxies with stellar masses above
$2.3 \cdot 10^{10} M_{\odot}$ with an initial mass of $4.6 \cdot
10^5 M_{\odot}$.

Four of our simulations are `test' runs with a
smaller box size of $(68 \mathrm{Mpc})^3$, which were performed to be able to
test the effect of the new black hole accretion and AGN feedback model
separately. The first run adopts the `original' black hole model as
described in \citet{Hirschmann} to which we refer as the fiducial
model. The second run adopts only the new accretion model (NAM), the
third run only adopts the new feedback model (NFM), and finally, our fourth run
combines both new implementations (NFAM). 

The other two simulations have the same resolution but a larger box
size of $(182 \mathrm{Mpc})^3$ to achieve a larger statistical sample
of galaxies and black holes. 
The first box uses the original implementation of black hole growth
and the second box adopts the NFAM model, enabling us 
to statistically see the effects of the new model, in particular on
the more massive galaxy and black hole population. 

As described in section \ref{Theoretical_model} in detail, the
NAM, NFM and NAFM models contain improvements of the black hole
model regarding the calculation of the accretion rate and/or the
feedback energy of black holes:  
\begin{enumerate}
\item{NAM: For the estimation of the black hole accretion rate we use
    different boost factors for cold ($\alpha=100$) and hot
    ($\alpha=10$) gas. For this run we use the fiducial feedback model.}
\item{NFM: For the calculation of the energy of the AGN feedback we
    consider not only radiative, but also mechanical feedback. 
The two different feedback mechanisms have different efficiencies. 
The radiative efficiency $\epsilon_\mathrm{r}$ depends on the black
hole mass and the Eddington ratio, whereas the outflow efficiency
$\epsilon_\mathrm{o}$ depends only on the Eddington ratio. 
Like in the fiducial model only a fraction $\epsilon_\mathrm{f}$ of
the radiation couples to the surrounding medium. 
Both kinds of feedback are implemented as thermal feedback. Hence, the
total feedback energy is computed with equation
(\ref{feedback_energy_new}).
We use the old accretion model for this simulation.
}
\item{NFAM: Our final run contains both the new feedback and the new accretion model. 
}
\end{enumerate}

The new feedback model as shown in Fig. \ref{feedback_model_plot} was
implemented into the code using equation (\ref{epsilon_r_final}) and
(\ref{epsilon_o_final}). In reality the slope $\beta$ can be
 between 0 and 1. However, the choice of $\beta$ does not
play a significant role for the simulations, as the mechanical outflow
dominates over the radiation in the radio regime. Furthermore, the 
AGN luminosities are not calculated during the simulation but only
for the analysis afterwards. Thus, we choose the fixed value of
$\beta=0.5$ for all simulations.  

For the NAM run and the two fiducial runs we use the standard feedback
model with $\epsilon_\mathrm{f}=0.15$ and a constant radiative efficiency
$\epsilon_\mathrm{r}=0.2$ \citep{Hirschmann}. In the other runs we use
$\epsilon_\mathrm{f}=0.2$. The parameters of the
simulations used in this work are summarized in Table
\ref{boxes_resolutions}.

Note that we identify the dark matter haloes and the
corresponding galaxies in the simulation using the friends-of-friends and then the SUBFIND algorithm
(\citealt{Dolag09}, \citealt{Springel_subfind}).

\section{Results}
\label{Results}

\subsection{Black hole growth}
\label{results_comparison}

\subsubsection{Black hole-galaxy mass scaling relations at $z=0$}
\begin{figure}
  \includegraphics[trim = 5mm 0mm 3mm 10mm, clip, width=0.5\textwidth]{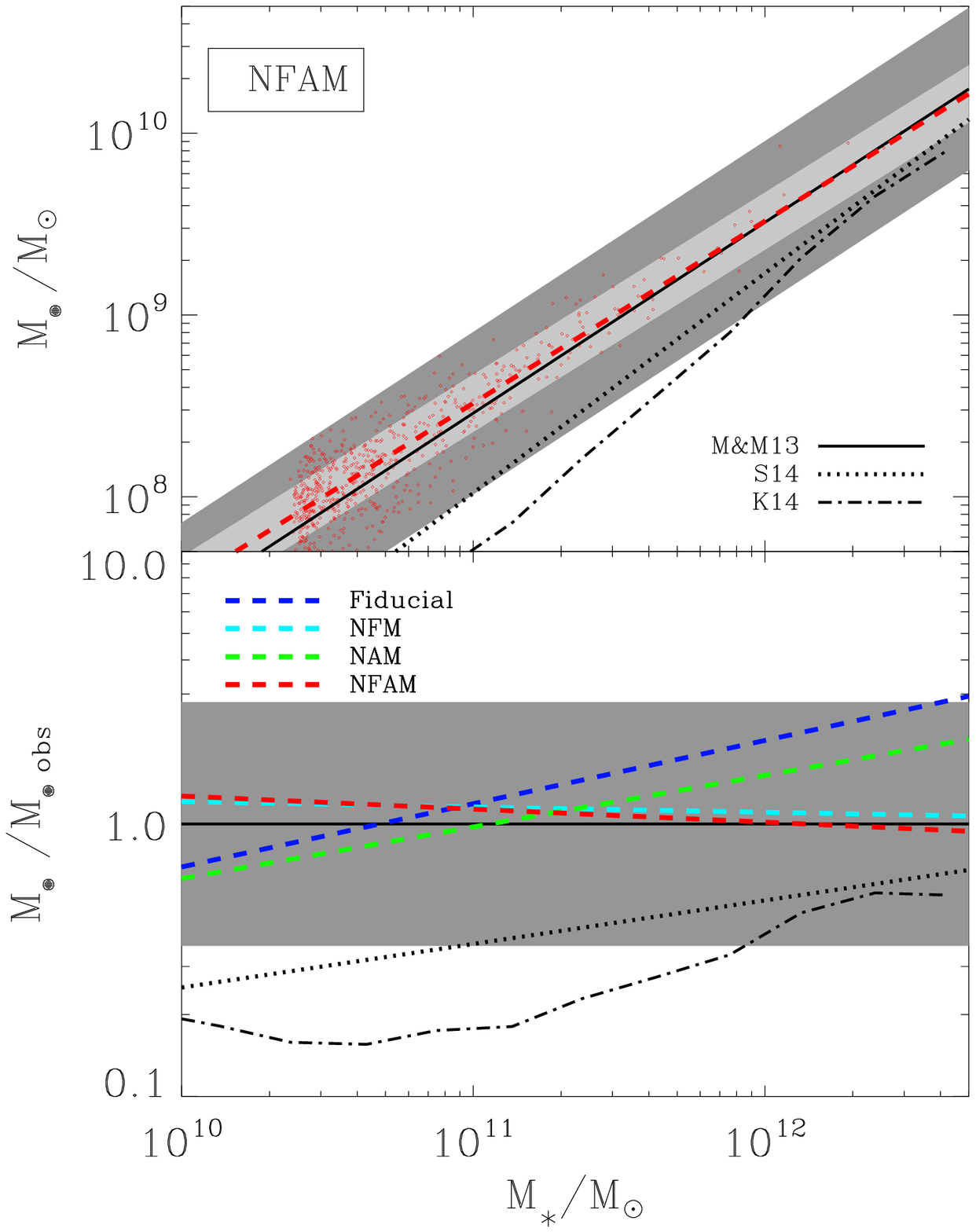}          
  \caption{Upper panel: present-day relation between the
    black hole mass and the host galaxy stellar mass for
      68Mpc/hr NFAM run. The dots represent the black holes in the
    simulations at $z=0$. The solid black line shows the fit to the
    observations by \citealt{McConnell} (M\&M13) and the dark shaded
    area the corresponding $1 \sigma$-error. The dashed lines
    illustrate the fit to our simulation for 
    $M_{\bullet}>5\cdot 10^7 M_{\odot}$ (to exclude seeding effects) and the
    light shaded area the corresponding $1 \sigma$-error.
    For comparison with other simulations we also show the
      results from \citealt{Sijacki_2014} (S14) and \citealt{Khandai}
      (K14) as dotted and dotted-dashed lines.
    Lower panel: Ratio of the simulated black hole mass in all
      different models (Fiducial: dark blue, NFM: light blue, NAM:
      green, NFAM: red) to the observed black hole mass $M_{\bullet
        \mathrm{obs}}$ (\citet{McConnell}, black solid line and grey
      shaded area) versus the galaxy stellar mass.  
    } 
\label{bh_bulge_feedback_model}
\end{figure}
\begin{table}
\centering
\begin{tabular} {|l|l|l|l|}
\hline
{} & $a$ & $b$ & $\sigma$\\
\hline
McConnell \& Ma (2013) & $8.46\pm 0.08$ & $1.05\pm 0.11$ & 0.45\\
68Mpc/hr fiducial model & 8.53 & 1.28 & 0.17\\
68Mpc/hr NFM & 8.52 & 1.03 & 0.16\\
68Mpc/hr NAM & 8.44 & 1.24 & 0.19\\
68Mpc/hr NFAM & 8.51 & 1.00 & 0.16\\
182Mpc/hr fiducial model & 8.46 & 0.93 & 0.15\\
182Mpc/hr NFAM & 8.40 & 1.09 & 0.14\\
\hline
\end{tabular}
\caption{Best-fit parameters and standard deviation for our runs in
  comparison to the observations by \citet{McConnell}. All black
  holes with masses smaller than $5\cdot 10^7 M_{\odot}$ have been
  excluded for the fit. For the 182Mpc/hr runs we took only stellar
  masses below $10^{12} M_{\odot}$ into account to exclude clusters.} 
\label{fit}
\end{table}
The upper panel in Fig. \ref{bh_bulge_feedback_model} shows
the predictions for the present-day $M_{\bullet}$-$M_*$ relation for the
68Mpc/hr NFAM simulation.
In our simulations $M_*$ is the total stellar mass of a galaxy and not
only the stellar mass of the bulge, because our resolution is not high
enough to resolve the internal structures of the individual galaxies. 
Hence, all galaxies consist mainly of a spheroidal component.
The solid black lines in Fig. \ref{bh_bulge_feedback_model} indicate
the observations of \citet{McConnell} 
and the dashed line is the fit for all black holes in our simulations
with $M_{\bullet}>5\cdot 10^7$. This threshold is necessary to
exclude newly seeded black holes, as they are seeded far below
the relation and need time to grow onto the relation.
Black holes with masses above  $M_{\bullet}>5\cdot 10^7$ are
close enough to the $M_{\bullet}$-$M_*$ relation to exclude seeding
effects. The figure shows the excellent agreement of our NFAM model 
with observations, in particular in comparison to other simulations,
i.e. the Illustris simulation \citep{Sijacki_2014} and the
MassiveBlack-II simulation \citep{Khandai}. 
The dark grey shaded area marks the $1 \sigma$-scatter of the
observations and the light grey shaded area the $1 \sigma$-scatter for
our simulation. For a quantitative comparison with the observations,
Table \ref{fit} shows the best-fitting parameters $a$ and $b$
corresponding to the fit function
$\mathrm{log}(M_{\bullet}/M_{\mathrm{\odot}}) = a + b \cdot
\mathrm{log}(M_{*}/10^{11} M_{\mathrm{\odot}})$ for all six runs.
It also contains the 
$1 \sigma$ scatter of \citet{McConnell} and our simulations. For the
182Mpc/hr runs, we consider only stellar masses below $10^{12}
M_{\odot}$ to exclude the central galaxies of very massive clusters
(see discussion in section \ref{results_evolution}).

While the slope of the $M_{\bullet}$-$M_*$ relation turns out to be
relatively insensitive to the values of $\epsilon_\mathrm{r}$ and
$\epsilon_\mathrm{f}$, the normalization depends strongly on these
parameters as already shown by \citet{DiMatteo05}, because the final
black hole mass follows the proportionality $M_{\bullet} \propto
(\epsilon_\mathrm{f} \epsilon_\mathrm{r})^{-1}$. Hence, many recent
simulations which include black holes (e.g. \citealt{DiMatteo05},
\citealt{Robertson}, \citealt{DeGraf_2010}, \citealt{Hirschmann})
tuned these parameters in order to reproduce the normalization of the
observed $M_{\bullet}$-$M_*$ relation. 
In addition, the normalization depends on the cooling function
\citep{Churazov}, i.e. the values of $\epsilon_\mathrm{r}$ and
$\epsilon_\mathrm{f}$ must be larger to get the same normalization
if the cooling is more effective.
Since $\epsilon_\mathrm{r}$ is not a constant parameter in our
new AGN feedback model, the slope of the $M_{\bullet}$-$M_*$
relation changes. This is shown in the lower panel of
Fig. \ref{bh_bulge_feedback_model}. 
Here we show the ratio of the simulated to the observed black hole
mass (from \citealt{McConnell}) versus the galaxy stellar mass for all
different models, i.e. the Fiducial, NFM, NAM and NFAM runs (colored
dashed lines), as well as for the results from \citet{Sijacki_2014} and
\citet{Khandai} (black dotted and dotted-dashed lines, respectively).   
Since they use a constant radiative efficiency, their slopes are
similar to our fiducial simulation. In our new feedback model,
however, $\epsilon_\mathrm{r}$ is not a free parameter anymore.
Therefore, it is encouraging that 
\textit{both the slope and the normalization of the
  $M_{\bullet}$-$M_*$ relation are self-consistently predicted with
  less free parameters than in the standard model.}

However, even in our new model one free parameter remains,
i.e. the fraction of radiation coupling to the surrounding medium
$\epsilon_\mathrm{f}$, for which we choose a value of
$\epsilon_\mathrm{f}=0.2$ (to be consistent with the observed
relation)\footnote{Note that this value depends on the resolution,
  because at lower resolutions the feedback energy is spread further
  away from the black hole. Hence, for our simulations, this value is
  comparatively high.}. For lower efficiencies the feedback would be
higher and the black holes would grow too much.
We would like to remark that the normalization of the
$M_{\bullet}$-$M_*$ relation in simulations always depends on the
observations used for the calibration of $\epsilon_\mathrm{f}$. 
However, there are discrepancies in observational estimations of the
$M_{\bullet}$-$M_*$ relation. 
For example, \citet{Scott} find a slightly higher normalization, but a
similar slope as \citet{McConnell}, which would change the
calibration of $\epsilon_\mathrm{f}$.  

In our simulations, the NFAM model reproduces the observed
slope better than the Fiducial model,
in which the black holes accrete slightly too much gas,
resulting in too large masses, particularly at low redshifts and in
the most massive galaxies. The new AGN feedback model is more
efficient in preventing gas accretion onto massive black holes. Thus,
the gas in the vicinity of the black hole has a higher thermal and
kinetic energy, which 
results in lower accretion rates. Consequently, as can be seen in
Fig. \ref{bh_bulge_feedback_model}, the massive end of the
$M_{\bullet}$-$M_*$ relation is now in excellent agreement with the
observations from \citet{McConnell}.  

Our second implementation is the separation of hot and cold gas (NAM).
For an increasing amount of hot gas in the vicinity of the black hole,
this results in slightly lower accretion rates due to the smaller
boost factor. Even if the new accretion model by itself cannot
prevent the most massive black holes from growing too much, it can
decrease the black hole masses slightly.
Consequently, a combination of both
modifications results in the best match with the observed
$M_{\bullet}$-$M_*$ relation. 

The best-fitting parameters in Table \ref{fit} summarize the excellent
agreement of the NFM-run and the NFAM-run with the observations.
Particularly, the slope $b$ is in better agreement with the observations
than in the other runs and also in the analysis of the Illustris
simulation shown by \citet{Sijacki_2014}. 
Note that in the simulations, the 1-$\sigma$ scatter is significantly
smaller than in the observations. As the typical measurement errors in
the observations are still substantial, future observations are needed
to distinguish, whether this relation indeed has such a small scatter as seen in
the simulations, or if there are still additional processes missing in
the simulations which influence the growth and evolution of the black 
holes.

Furthermore, the scatter in the black hole mass in the simulations
decreases with increasing black hole mass. 
This is most likely a consequence of statistical merging
(\citealt{Peng}, \citealt{Hirschmann_2010}, \citealt{Jahnke}) and is
also visible in the Illustis simulation \citep{Sijacki_2014}. 
Nevertheless, the relative role of AGN feedback and stasticial merging
in establishing the $M_{\bullet}$-$M_*$ relation and producing the
observed slope still remains a matter of debate. 

\begin{figure}
  \includegraphics[trim = 0mm 0mm 0mm 0mm, clip, width=0.5\textwidth]{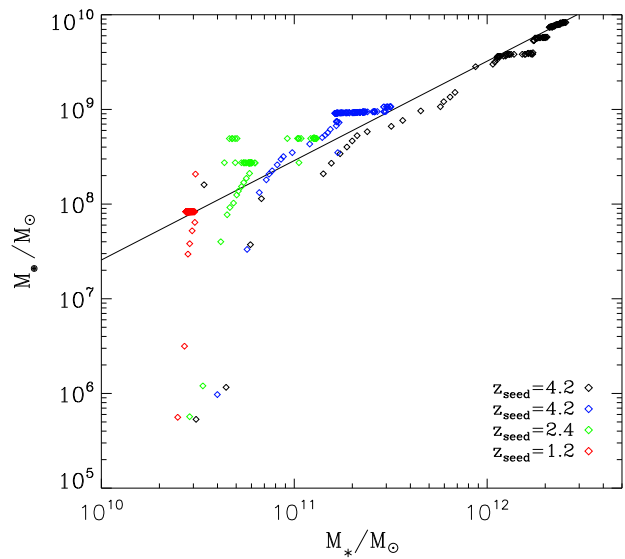} 
  \caption{Evolution of the total black hole mass and the
    corresponding host galaxy stellar mass of four haloes (diamonds in
    different colors) in the 68Mpc/hr NFAM simulation.
    The black line shows the fit from \citet{McConnell}
    }
  \label{bh_bulge_evolution_feedback_model}
\end{figure}

To explore black hole growth in our simulations in more detail,
Fig. \ref{bh_bulge_evolution_feedback_model} shows the cosmic
evolution of four black holes selected due to their different present-day
mass (different colors) on the $M_{\bullet}$-$M_*$
relation\footnote{The two outliers (black and red diamond with
  $M_\bullet \approx 2\cdot 10^8 M_{*}$) are due to temporary
  attributions to different haloes.}.
When black holes are merging,
the most massive progenitor is followed back in time. As can be seen
in this figure we can distinguish between two different phases of
black hole growth: during the first phase, they grow rapidly until
they reach the $M_{\bullet}$-$M_*$ relation and thus the Eddington
limit. In this phase black hole accretion is primarily triggered by
smooth accretion of cold gas, because below the Eddington limit AGN
feedback is not strong enough to suppress gas cooling.
Hence, the cold gas reservoir is large enough to trigger black hole
growth. In our simulations, this phase is a consequence of the small
black hole seeding mass. 
However, recent observations seem to indicate that the slope of the
$M_{\bullet}$-$M_*$ relation is steeper for black holes with masses
below $10^8 M_{\odot}$ (\citealt{Graham}, \citealt{Scott}).
Therefore, we can speculate that the phase of rapid black hole growth
is actually present and that simulations in which black holes are
seeded on or above the $M_{\bullet}$-$M_*$ relation might miss the
first phase of black hole growth.
  
In the second phase black holes grow along the $M_{\bullet}$-$M_*$ relation.
In this phase, gas cooling and AGN feedback are in equilibrium and
hence both star formation and black hole growth are suppressed. 
Only the in-fall of cold gas either in the form of streams or clumps
as well as merger events can trigger star formation and black hole
growth during this period. 

To demonstrate that at low redshifts black holes grow faster
compared to the growth of the stellar mass than at high redshifts,
we show exemplarily the results for four typical objects, where we
verified that they reflect the typical growth of BHs with the 
chosen final mass. 
For example, the stellar mass of the host galaxy corresponding to the
red diamonds grows very little, whereas the black hole mass increases
by more than two orders of magnitude. This galaxy reaches the
  $M_{\bullet}$-$M_*$ relation already 1.08 Gyr after the seeding.
In contrast, the stellar mass of the host galaxy corresponding to the
black and blue diamonds grows much more during the first phase of
black hole growth. 
Here, the object reaches\footnote{We excluded the outlier (black diamond on
the left with $M_\bullet \approx 2\cdot 10^8 M_{*}$).} 
the $M_{\bullet}$-$M_*$ relation after 2.29 Gyr.
This trend is also visible in Fig. \ref{bh_bulge_feedback_model_all},
which shows the $M_{\bullet}$-$M_*$ relation at different redshifts,
in particular when looking at the data points corresponding to the
lowest stellar masses. The figure will be discussed later in more detail. 
Hence, we suspect that the black hole mass at the threshold between
the two phases -- namely when the $M_{\bullet}$-$M_*$ relation is
reached -- depends on the seeding redshift.  
We suggest, that these differences might be a consequence of the star
formation rate, which decreases with time (see section
\ref{SFR_evolution}). 

Furthermore, since black holes are seeded upon a certain galaxy mass,
they are seeded earlier in a dense environment and can thus become
more massive. 
We plan to study the evolution of black holes and their host galaxies
in a forthcoming study in more detail, performing a simulation with
resolution high enough to resolve the internal structure of galaxies. 
In particular, we are interested in the effect of merger events on
black hole growth and star formation, 
because the black hole and stellar masses in
Fig. \ref{bh_bulge_evolution_feedback_model} seem to grow mainly in
steps after reaching the $M_{\bullet}$-$M_*$ relation. 
These steps also explain the scatter around the $M_{\bullet}$-$M_*$
relation in our simulations. 
It furthermore indicates, that black hole growth and star formation
are both triggered by merger events. However, for this study it is
more important to increase the box size instead of the resolution, in
particular to extend our simulation results towards more massive
galaxies and black holes. 

\subsubsection{Evolution of the black hole mass function}
\begin{figure}
  \includegraphics[trim = 5mm 10mm 20mm 120mm, clip,
  width=0.5\textwidth]{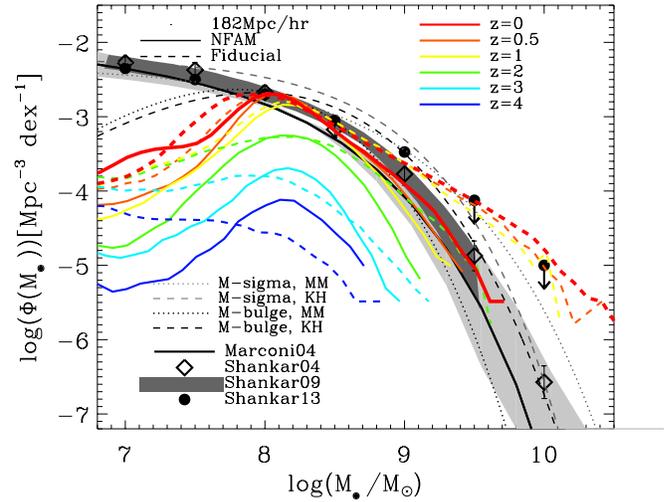} 
  \caption{
    Black hole mass function of the fiducial (dashed coloured lines) and
    the NFAM (solid coloured lines) 182Mpc/hr simulation at different
    redshifts. 
    For comparison we show observations from \citealt{Marconi} (black solid line),
    \citealt{Shankar04} (black diamonds and lines with grey shaded areas),
    \citealt{Shankar09} (dark grey shaded area)
    and \citealt{Shankar13} (black dots).
    To show the uncertainties in deriving black hole mass functions from
    observations,
    we show as dotted and dashed grey curves the black hole mass functions derived from the best fit velocity dispersion function and stellar mass function
    from \citet{Bernardi_2010} using different scaling relations, i.e. from \citealt{McConnell} (MM) and \citealt{Kormendy} (KK). 
    } 
\label{BMF}
\end{figure}
Fig. \ref{BMF} shows the black hole mass function of both the fiducial
and the NFAM 182Mpc/hr run.
We compare our simulations to observed black hole mass
  functions of the local universe by \citet{Marconi},
  \citet{Shankar04}, \citet{Shankar09} and \citet{Shankar13}.  We
  would like to remark that the uncertainties in these relations are
  large, in particular because the black hole masses are estimated
  using different scaling relations as recently discussed by
  \citet{Shankar13} and therefore, 
  we also show the black hole mass functions derived from
  the best fit velocity dispersion function and stellar mass function from 
   \citet{Bernardi_2010} using different scaling relations, i.e. from \citealt{McConnell} (dotted grey lines) and \citealt{Kormendy} (dashed grey lines).
  Since the high mass end of all of these curves is lower than in \citet{Shankar13},
   we take -- following their discussion --
   the two data points at the high mass end of \citet{Shankar13}
  as upper limits. One should also keep in mind that as discussed in
  \citet{Tundo}, the different black hole scaling relations are not
  necessarily consistent with each other or with the
  $M_{\bullet}$-$M_*$ relation from \cite{McConnell}, which we use in
  this work to calibrate the value of the free parameter
  $\epsilon_f$. The uncertainties in the scaling relations are also
  reviewed and discussed in \citealt{Kormendy}. 

The high mass end of the fiducial simulation is just in agreement with the upper
  limits of \citet{Shankar13}, but the NFAM simulation matches
  previously published black hole mass functions much better, because
  the new accretion and feedback models suppress the growth of massive
  black holes more efficiently. As already shown in
  Fig. \ref{bh_bulge_feedback_model}, the smaller masses of the most
  massive black holes are mainly caused by the new feedback scheme,
  where  the mass dependency of the radiative efficiency for the model
  is taken from \citet{Davis}, which is quite similar to the results
  presented in \citet{Trakhtenbrot}. From a theoretical point of view,
  this relation is motivated by the fact that the spin of the black
  hole should increase with mass. However, the slope of this relation
  might actually be flatter than in \citet{Davis} due to selection
  effects (see discussion in \citealt{Raimundo} and \citealt{Laor}). 
  Thus, the massive end of the black hole mass function of the NFAM
  simulation could be a lower limit. Furthermore, we already mentioned
  that it is uncertain whether in general the normalization of the
  $M_{\bullet}$-$M_*$ relation could be larger than in
  \citet{McConnell}. 

For less massive galaxies, the effects of the seeding become
  dominant which cause the deviation from the observed black hole
  mass function at small masses. However, especially at low masses,
  observations are uncertain and only give an upper limit
  \citep{Shankar13}, in particular because pseudo-bulges do probably
  not follow the observed scaling relations like the
  $M_{\bullet}-\sigma$ relation or the $M_{\bullet}$-$M_*$ relation as
  reviewed by \citet{Kormendy}.

\subsubsection{Evolution of the black hole-galaxy mass scaling relations}
\begin{figure}
  \includegraphics[trim = 0mm 0mm 0mm 0mm, clip, width=0.5\textwidth]{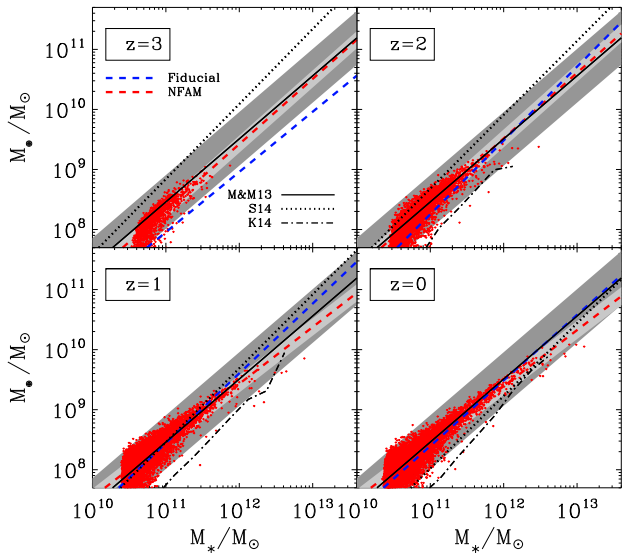} 
  \caption{Evolution of the relation between the black hole mass and
    the host galaxy stellar mass for the NFAM 182Mpc/hr run (red
    dots). The dashed lines are fits for both 182Mpc/hr runs including
    all black holes with masses larger than $5\cdot 10^7 M_{\odot}$
    and stellar masses with masses smaller than $10^{12} M_{\odot}$ to
    exclude clusters.
    The light grey shaded area marks the corresponding $1 \sigma$-error of the NFAM run.
    The black line with the dark grey shaded area represents the fit through the
    observations from \citet{McConnell} with the $1 \sigma$-error.
    The dotted and dotted-dashed lines show the results from
    other simulations, i.e. from \citet{Sijacki_2014} and
    \citet{Khandai}.
    }  
  \label{bh_bulge_feedback_model_all}
\end{figure}

Fig. \ref{bh_bulge_feedback_model_all} shows the relation between the
black hole mass and the stellar mass of the host galaxy for our NFAM
182Mpc/hr run at different redshifts, again in comparison to the
observations by \citet{McConnell} and the simulations from
\citet{Sijacki_2014} and \citet{Khandai}. Again, we only show black
holes with masses above $5\cdot 10^7 M_{\odot}$. Below this limit
black holes generally grow fast,
while $M_*$ stays relatively constant until they reach the
$M_{\bullet}$-$M_*$ relation. The reason is the equilibrium
between AGN feedback and gas cooling, when black holes accrete with
$\dot{M}_{\mathrm{Edd}}$ as described by \citet{Churazov}. Afterwards
black holes can only grow along the $M_{\bullet}$-$M_*$ relation
together with their host galaxy through smooth accretion or merging.

In the NFAM run, the $M_{\bullet}$-$M_*$
relation is much earlier in place than in the original run, namely
already at $z=3$.
Furthermore, the panels at $z=2$ and $z=1$ show that
in the fiducial simulation the slope of the $M_{\bullet}$-$M_*$
relation is larger than at $z=0$, where it is in agreement with the
observed $M_{\bullet}$-$M_*$ relation. 

In our very massive galaxies ($M_*\approx10^{13}M_\odot$),
i.e. the central galaxies of galaxy clusters,
most black holes are lying slightly below the $M_{\bullet}$-$M_*$
relation. This is most likely caused by a still too large stellar
mass in these very massive galaxies, also visible in the high mass
excess of the stellar mass function and the still too large baryon
conversion efficiency for large haloes as discussed later on. 
The reason for the overestimation of stellar masses of cluster
galaxies might be the purely thermal feedback in our model, which
fails to reproduce the mechanical feedback in such massive systems,
visible as large X-ray cavities in observed clusters.
Hence, an implementation of mechanical jets
(e.g. \citealt{Ostriker}, \citealt{Dubois_zoom}, \citealt{Choi_2014}) might play an important role for future
simulations, in which both the resolution and the size of the
cosmological boxes will get larger and larger.
Furthermore, in our analysis we do not distinguish between
the stars belonging to the central galaxy and the ones which would be
related to the intra cluster light (ICL), which can be substantial for
such massive systems. It is also possible that some merging systems
are identified as one galaxy. Thus, the predicted
stellar mass for cluster galaxies might actually be slightly larger
than in observations. 

For comparison, Fig. \ref{bh_bulge_feedback_model_all} also includes
the fit to the data points of the fiducial model, where black holes in
galaxy clusters are substantially more massive compared to the stellar
mass, especially at redshifts around $z=1$. Although the fit at $z=0$
is in agreement with the fit from \citet{McConnell}, it is evident from
the black hole mass function that the black hole masses are too large
at the high-mass end implying that the galaxy stellar masses must be
too large (compensating for the large black hole masses) which will be
investigated in more detail in section \ref{results_evolution}. 

\subsubsection{Eddington ratio distribution}
\begin{figure}
  \includegraphics[width=0.5\textwidth]{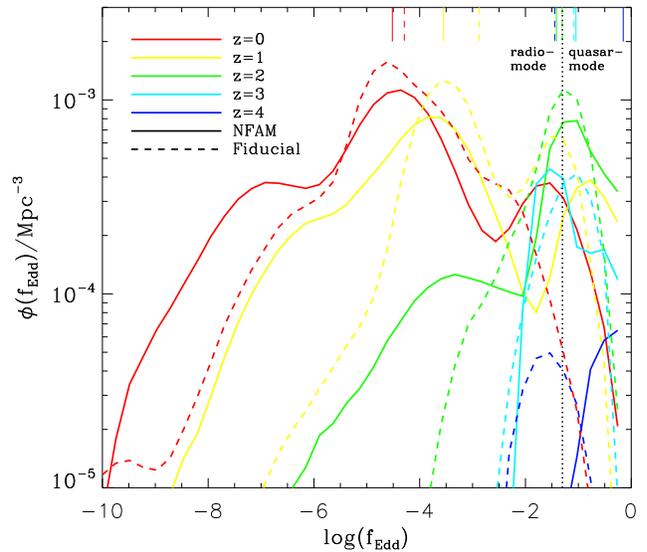}
  \caption{Eddington ratio distributions for the two 182Mpc/hr
    simulations at different redshifts. 
The black dotted vertical line marks the threshold between radio-mode
and quasar-mode. 
The vertical lines in the top show the mean values.
}
\label{fedd}
\end{figure}
The modifications in our NFAM simulations are also expected to
significantly affect the Eddington ratios of the black holes. 
Therefore, in Fig. \ref{fedd} we present the Eddington ratio
distributions of both 182Mpc/hr simulations at different redshifts. 
The black dotted vertical line shows the threshold between radio-mode
and quasar-mode and the vertical lines in the top mark the mean
values. 
For redshifts below $z=3$ the Eddington ratios are clearly smaller in
the NFAM run than in the fiducial simulation. 
For higher redshifts the Eddington ratios in the NFAM run are larger
than in the fiducial simulation. 
We suggest that the wide range of values for the feedback efficiency
leads to broader distributions. 
Especially the range of very low accretion rates is represented much
better in the NFAM simulation than in the fiducial run. 

In contrast to the recent study from \citet{Sijacki_2014} our
simulations -- in particular the NFAM run -- show two peaks in the
Eddington ratio distribution for $z<4$, one in the radio-mode and a
second peak either in the radio-mode or in the quasar-mode. 
This indicates that we have a clear separation between two accretion modes.
In the fiducial model, where a step function was used to distinguish
between radio-mode and quasar-mode \citep{Hirschmann}, the two peaks
are only visible at $z=1$. 
In the NFAM simulation, the second peak appears at $z=3$ in the
quasar-mode. For smaller redshifts it is much more distinct. 
Interestingly, at $z=1$ and $z=2$, which is the redshift range where
most quasars are observed, a very clear second peak is visible in the
quasar-mode. 
For $z=4$ the Eddington ratios are even higher, because here the first
phase of black hole growth is dominant. 
At $z=0$ both peaks are in the radio-mode and even a third peak is
visible at very low Eddington ratios. 

\subsection{Evolution of the stellar mass function}
\label{results_evolution}
\begin{figure*}
  \includegraphics[trim = 0mm 0mm 9.5mm 0mm, clip,
  width=\textwidth]{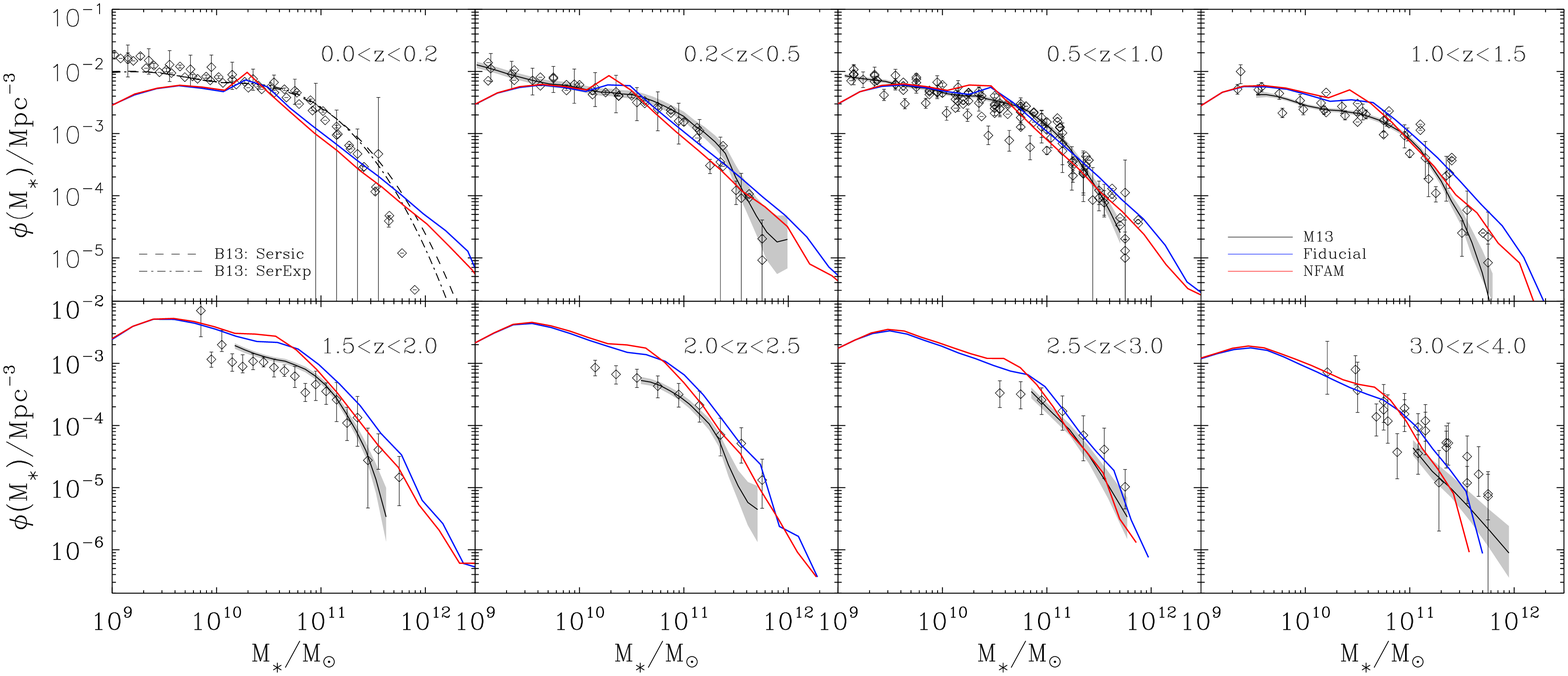} 
  \caption{Stellar mass functions in different redshift ranges
      for the fiducial (blue lines) and the NFAM (red lines) 182Mpc/hr
      runs. The solid black lines with the shaded areas show
    the observed stellar mass functions presented by \citealt{Muzzin}
    (M13) and their Poisson errors. The black diamonds are
    observations from \citet{Panter}, \citet{Cole}, \citet{Bell},
    \citet{Perez_Gonzalez}, \citet{Borch}, \citet{Bundy}, \citet{Drory},
    \citet{Fontana} and \citet{Marchesini}.
    The black dashed and dotted-dashed lines show the
    result from \citealt{Bernardi} (B13) using a Sersic model and a
    Sersic-bulge + exponential-disc model.}  
  \label{SMF}
\end{figure*}
\begin{figure*}
  \includegraphics[trim = 0mm 0mm 9.5mm 0mm, clip, width=\textwidth]{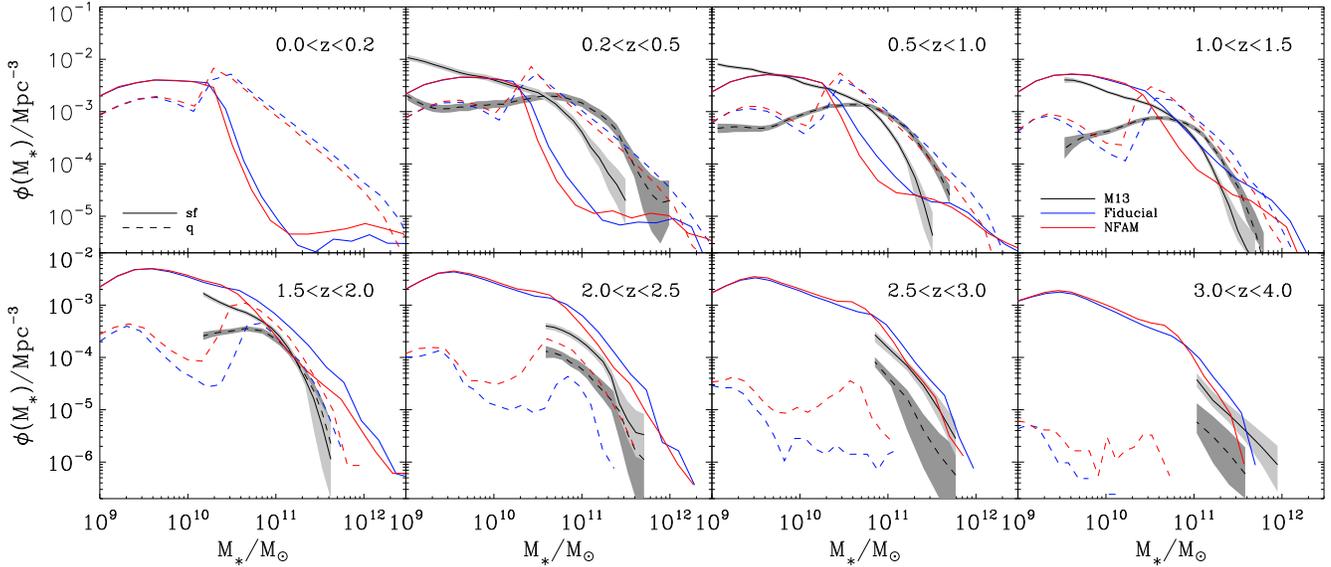}
  \caption{Stellar mass functions of quiescent (dashed lines) and
    star-forming (solid lines) galaxies in different redshift ranges
    for the fiducial (blue lines) and the NFAM (red lines)
      182Mpc/hr runs. For the threshold between quiescent and
    star-forming galaxies we use the specific star formation rate of
    $0.3 / t_{\mathrm{ Hubble}}$ following Franx et al. (2008). The
    black lines with the shaded areas (light grey for star forming and
    dark grey for quiescent galaxies) show the observations from
    \citealt{Muzzin} (M13) and their Poisson errors.} 
  \label{SMF_sf_q}
\end{figure*}
Fig. \ref{SMF} shows the evolution of the stellar mass function in the
simulations (blue: fiducial model, red: NFAM model) and observations
(black symbols from \citealt{Panter}, \citealt{Cole}, \citealt{Bell},
\citealt{Perez_Gonzalez}, \citealt{Borch}, \citealt{Bundy},
\citealt{Drory}, \citealt{Fontana} and \citealt{Marchesini} and black
lines from \citealt{Muzzin} and \citealt{Bernardi}).
The figure illustrates that the new
feedback scheme can slightly suppress late star formation at the
high-mass end, mainly because the radiative efficiency now depends on
the black hole mass. Hence, compared to the fiducial model, the
modifications in the NFAM model lower the amount of massive galaxies
resulting in an overall better match with the massive end of the
observed SMF, at least down to $z=0.2$.  

For the entire redshift range, a small peak in the SMFs is
visible at stellar masses of about $2\cdot 10^{10} M_{\odot}$.
The origin of this peak is caused by a subtle effect of our black hole
seeding. Since black holes are seeded below the $M_{\bullet}$-$M_*$
relation, the AGN feedback is efficient during the first phase of black
hole growth and hence suppresses star formation until the equilibrium
between cooling and AGN feedback is reached.
During that phase, the stellar mass stops growing and consequently,
there are more galaxies with a certain stellar mass.
The peak moves towards higher stellar masses at higher redshifts
because of the effect seen in
Fig. \ref{bh_bulge_evolution_feedback_model}, namely that black holes
which are seeded earlier have larger stellar masses when they reach
the $M_{\bullet}$-$M_*$ relation.

The overestimation of the low-mass end of the stellar mass function at
high redshifts happens  most likely due to the chosen wind model
(constant winds as in \citealt{Springel_Hernquist}) as described by
\citet{Hirschmann} in more detail. Apart from that, our simulations -
especially the NFAM run - are in good agreement with observations at  
high redshifts. 

For $z<0.2$, the high-mass end is still overestimated. However, we
have to keep in mind that observations in this mass range contain also
relatively large uncertainties. \citet{Bernardi} showed that different
measurements of stellar masses differ from each other significantly,
especially at the high-mass end. They demonstrate that the stellar
masses are higher using a Sersic model instead of standard
models. Their fits using a single Sersic and a Sersic-bulge +
exponential-disc model are shown as black dashed and dotted dashed line  
in the upper left panel of Fig. \ref{SMF}. In comparison to other
observational estimates this is in better agreement with our
simulations. Nevertheless, the high-mass end still appears to be
slightly overestimated in our simulations as also indicated by the
massive end of the $M_{\bullet}$-$M_*$ relation
(see lower right panel of Fig. \ref{bh_bulge_feedback_model_all}).

To study the effect of our new accretion and feedback models on the
stellar masses in more detail, Fig. \ref{SMF_sf_q} shows the stellar
mass functions separately for quiescent and star-forming galaxies in
our simulations -- again in comparison to the observations from
\citet{Muzzin}. Following \citet{Franx} we use a specific star formation
rate of $0.3 / t_{\mathrm{ Hubble}}$ as threshold to distinguish
between quiescent and star-forming galaxies.
We would like to mention that this is a different selection
criterion than in the observations, where a threshold in the UVJ
diagram is used \citep{Muzzin}. Hence, this criterion might lead to
discrepancies with the observations,
which may e.g. falsely identify metal-rich, star-forming galaxies
to be red and thus quiescent.

Fig. \ref{SMF_sf_q} illustrates that our new implementations increase
the amount of quiescent galaxies at $z>1.5$. Consequently, for this
redshift range, the discrepancies between simulated and observed SMFs are much smaller for the NFAM simulation than for the Fiducial run.
Star formation is suppressed, when
cooling and AGN feedback are in equilibrium \citep{Churazov} and the
gas in the vicinity of the AGN cannot cool enough to form stars. 
Hence, the increase of the amount of quiescent galaxies can be
explained with the upper left panel in
Fig. \ref{bh_bulge_feedback_model_all}, which shows that the
$M_{\bullet}$-$M_*$ relation -- and thus the phase of equilibrium --
is earlier in place for the NFAM run. 
This is due to higher black hole accretion rates during the phase of
rapid black hole growth as a consequence of both new implementations:
firstly, the new accretion model leads to higher accretion rates when
cold gas dominates. Secondly, the new feedback model results in less
AGN feedback for low black hole masses and thus to lower gas
temperatures. 

In contrast to the equilibrium phase, which can be associated with the
radio-mode, the phase of star formation and rapid black hole growth is
not much affected by our new implementations. 
We conclude that the overestimation of the high-mass end is
mainly due to star-forming galaxies. At $z < 1$ the amount of
star-forming galaxies is too low for $2\cdot 10^{10} M_{\odot} < M_*
< 2\cdot 10^{11} M_{\odot}$. Firstly, this is an effect of the low
seeding mass of black holes, which also leads to the overproduction
of quiescent galaxies. Secondly, it is a consequence of the
overestimation of the high-mass end.

For both runs, Fig. \ref{SMF_sf_q} shows an artefact at low redshifts,
namely that the amount of star-forming galaxies decreases rapidly
after the seeding of black holes.
We speculate that this decrease might be due to our very low black
hole seeding mass, which leads to artificially high accretion rates. 
This also explains why the number of star-forming galaxies is reduced 
in the NFAM model compared to the fiducial one. 
Fig. \ref{bh_bulge_evolution_feedback_model} illustrates why this
artefact becomes even larger with decreasing redshift: 
for black holes that are seeded later, the evolutionary track during
the first phase of black hole growth is steeper then for early black
hole seeds. All in all Fig. \ref{SMF_sf_q} shows that our new
implementations cannot significantly improve the stellar mass
functions at low redshifts, but at high redshifts they predict a
larger amount of quiescent galaxies, which is in better agreement with
observations.    
 
\begin{figure}
  \includegraphics[trim = 0mm 6mm 0mm 13mm, clip, width=0.5\textwidth]{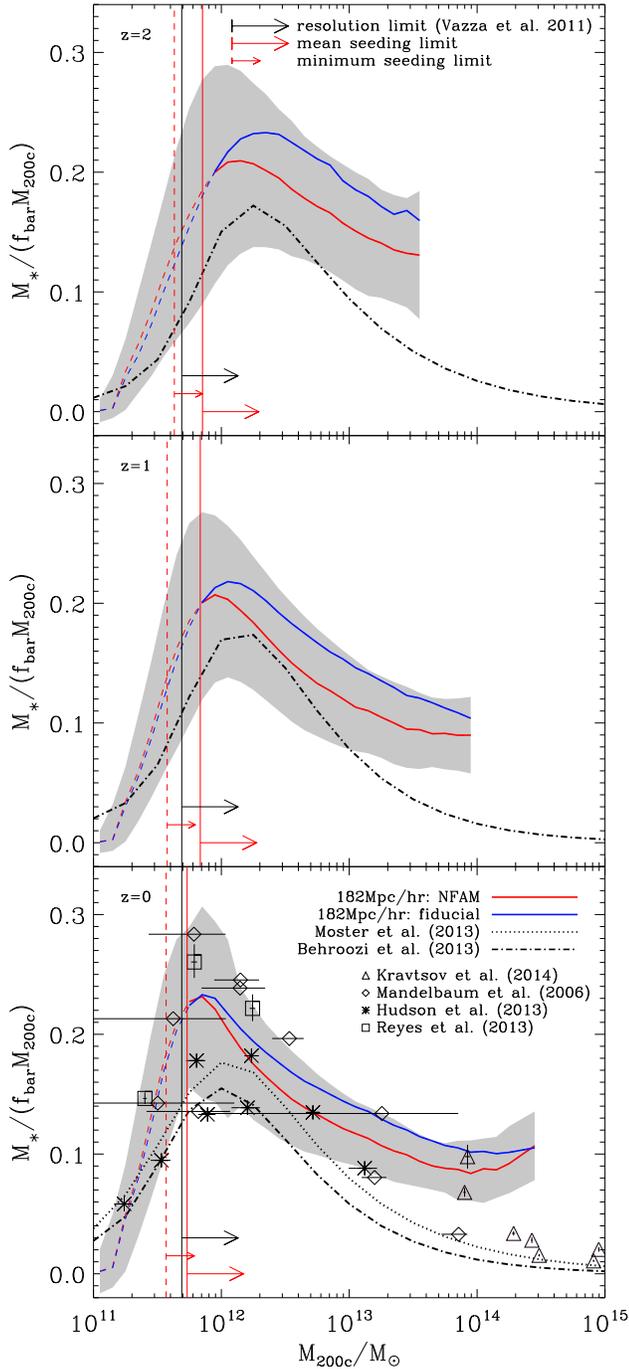}
  \caption{Mean baryon conversion efficiencies versus halo mass at
    different redshifts for the two 182Mpc/hr runs.
    The grey shaded area shows the $1 \sigma$-error of the
    NFAM run.
    The dashed and solid red vertical lines mark the minimum and mean
    value of $M_\mathrm{200c}$ in the NFAM simulation 
    corresponding to the minimum stellar mass for black hole seeds.
    Below the mean seeding limit our resolution does not allow
    reliable predictions (dashed lines). 
    The black vertical line shows the resolution limit for the baryon
    content as estimated by \citet{Vazza}, 
    which is given by 500 dark matter particles.
    We compare our simulation with
    abundance matching models (\citealt{Moster}, \citealt{Behroozi})
    and with observations estimating the halo mass with weak lensing
    (\citealt{Mandelbaum}, 
    \citealt{Hudson}, \citealt{Reyes}) or X-ray temperatures \citep{Kravtsov}.}
\label{BCE}
\end{figure}

To quantify how efficient baryons are converted into stars for a given
halo mass, we calculate the mean baryon conversion efficiencies, which
are defined as $M_*/(f_{\mathrm{bar}}M_{\mathrm{halo}})$, where
$f_{\mathrm{bar}}=0.17$ is the baryon fraction of the universe, for
different redshifts.
To be comparable to other studies we do not use $M_\mathrm{vir}$ for
the halo mass, but $M_\mathrm{200c}$, 
which is the mass inside the radius where the density is 200 times
larger than the critical density of the universe.
Fig. \ref{BCE} shows the conversion efficiencies versus halo mass for our
two 182Mpc/hr runs (different panels illustrate $z=0,1,2$).
The black vertical line shows the resolution limit for the baryon
content as estimated by \citet{Vazza}, 
which is given by 500 dark matter particles.  
Furthermore, the dashed and solid red vertical lines mark the minimum
and mean value of $M_\mathrm{200c}$, respectively, in the NFAM
simulation 
corresponding to the minimum stellar mass for black hole seeds.
Below the mean seeding limit our resolution does not allow reliable
predictions (dashed lines). 
The figure clearly shows, that the new implementations lower the
stellar content in a halo for a given mass above this limit, which is
also reflected by the reduced high-mass end of the stellar mass
functions (see Fig. \ref{SMF}). 
At $z=2$ and $z=1$, this effect is even stronger than at $z=0$.
The dotted and dotted-dashed black lines show the predictions of the
abundance matching models by \citet{Moster} and \citet{Behroozi}.
The peak at $M_{\mathrm{halo}} \approx 10^{12} M_{\odot}$ is in
agreement with these models, 
which also find a maximum baryon conversion efficiency of around 20
per cent. 
At larger halo masses, the stellar content decreases due to AGN
feedback and because the gas is consumed by star formation. 
Although the baryon conversion efficiencies in the NFAM simulation are
smaller than in the fiducial run, they are still higher
than in the abundance matching models of \citet{Moster} and
\citet{Behroozi} for $M_\mathrm{200c} > 10^{13} M_{\odot}$ galaxies. 

\begin{figure*}
  \includegraphics[width=\textwidth]{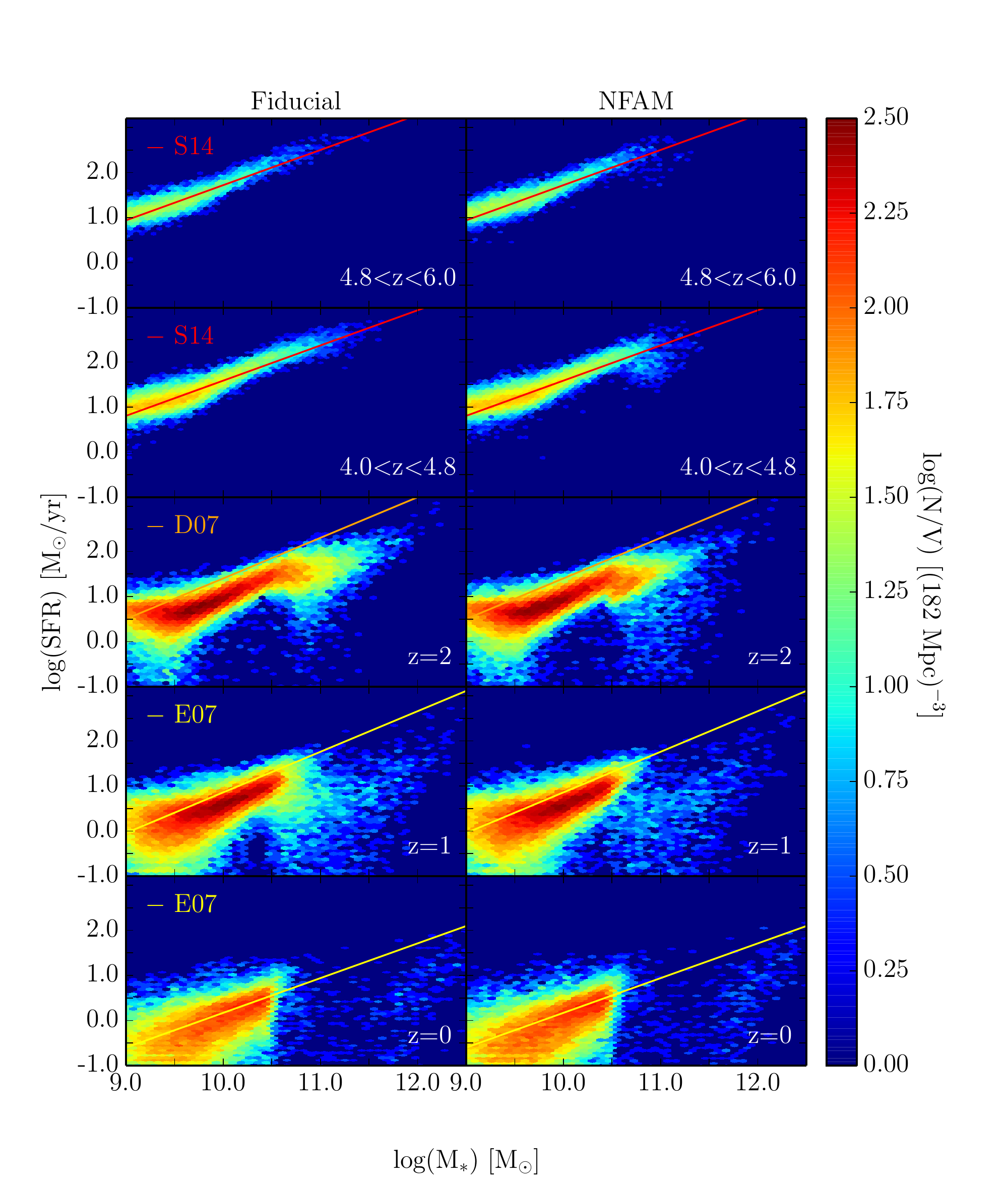}
\caption{Comparison of the star formation rates of all galaxies in the
  two 182Mpc/hr runs at different redshifts. The solid lines represent
  the observed main sequence of galaxies derived by \citealt{Steinhardt14}
  (S14), \citealt{Daddi} (D07) and \citealt{Elbaz} (E07).
}  
\label{SFR}
\end{figure*}

For the NFAM simulation, at low redshifts a slight ``upturn''
  of the baryon conversion efficiencies occurs for stellar masses
  above $10^{14} M_{\odot}$ corresponding to galaxy clusters due to
  too inefficient AGN feedback.
  This might indicate that other AGN feedback
  processes like mechanical jets should be included in future
  simulations.
  Since the most massive black holes accrete less in the
  NFAM model we suspect that there is more cold gas left to form stars
  than in the fiducial run. Therefore, the upturn is only visible
  in the NFAM simulation.   
However, except for the high-mass end, our simulations -- in
particular the NFAM run -- are in agreement with observations using
weak lensing (\citet{Mandelbaum}, \citet{Reyes} and \citet{Hudson}) or
X-ray temperatures \citet{Kravtsov} to estimate the total halo mass. 
\footnote{For the observations we computed $M_\mathrm{200c}$ out of
  $M_\mathrm{500c}$ using the NFW profile.} 

\subsection{Evolution of the star formation rate}
\label{SFR_evolution}
Fig. \ref{SFR} shows the SFR-stellar mass plane (number density is
color-coded) for our two 182Mpc/hr runs at different redshifts. 
The panels illustrate all galaxies classified as subhaloes
using the SUBFIND algorithm (\citealt{Dolag09},
\citealt{Springel_subfind}).
For comparison with observations, we also show the main sequence for
star-forming galaxies estimated by \citet{Steinhardt14} for $4<z<6$
(red line), by \citet{Daddi} for $z=2$ (orange line) and by
\citet{Elbaz} for $z=1$ and $z=0$ (yellow line). At $z=2$ and $z=1$,
the simulated SFRs at a given stellar mass are slightly below the
observations. 
This trend is also visible in the recently published analysis
of the Illustris simulation by \citet{Sparre}.
At $z=0$ and at redshifts above $z=4$ our simulation results are in
very good agreement with the observed main sequence,
independent of the adopted black hole model.
The redshift evolution of the SFR-stellar mass plane nicely demonstrates
that the most massive galaxies become more and more quiescent with
cosmic time. Furthermore, in the NFAM simulation star formation is
suppressed earlier than in the fiducial one.
This is consistent with Fig. \ref{SMF_sf_q}, where we demonstrated
that in the NFAM run the amount of quiescent galaxies is larger at
earlier times.  
In the NFAM simulation, the SFRs of the most massive galaxies decrease
already at redshifts above $z=4.8$ such that they lie below the
observed main sequence of star forming galaxies. In the fiducial
simulation, this decrease starts at redshifts below $z=4$. 
This may be unrealistic, because -- as shown in
Fig. \ref{SMF_sf_q} -- \cite{Muzzin} observe much more quiescent
galaxies at high redshifts ($z>3$) than in our fiducial simulation. 
Looking at the star formation main sequence of the Illustris
simulation \citep{Sparre} shows that this is not only a problem in our
fiducial run, but seems to be a general issue.  
Therefore, it is encouraging that in the NFAM run galaxies become
quiescent much earlier due to
both of our new implementations,
even if there are still discrepancies between the observed and simulated
SMFs for star-forming and quiescent galaxies. 
The new feedback model leads to a lower feedback energy for low black
hole masses, whereas for large black hole masses the AGN feedback is
stronger as long as the black holes are accreting in the quasar-mode
and star formation is suppressed. 

The new accretion model leads to lower accretion rates when the hot
gas phase dominates. Hence, black holes grow less strongly and the SFR
decreases already in less massive galaxies as can be seen in the
panels corresponding to $z=1$. From the earlier and more rapid
decrease of the SFR follows that at $z=1$ star-forming galaxies with
stellar masses above $2 \cdot 10^{10} M_{\odot}$ are more concentrated
along the observed main sequence in the NFAM simulation than in the
fiducial one. At $z=0$ there are only very few star-forming galaxies
above log$(M_*/M_{\odot})=10.5$, which is the mass at which AGN
feedback becomes important. At that redshift both runs predict
galaxies with similar SFRs at a given stellar mass. Hence, our
modifications mainly affect the evolution of high redshift galaxies. 

\begin{figure}
  \includegraphics[trim = 5mm 3mm 0mm 10mm, clip, width=0.5\textwidth]{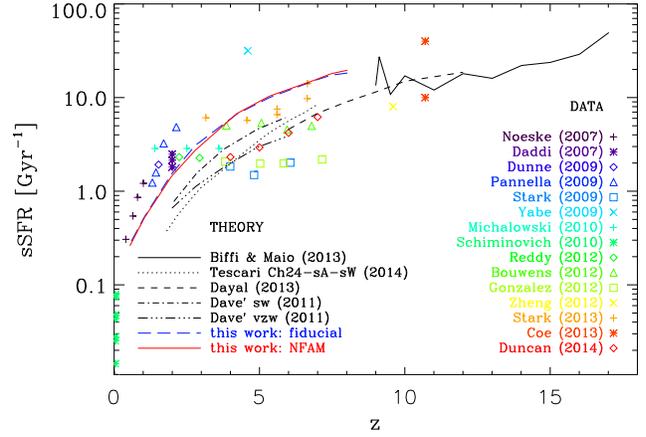}
\caption{History of the specific star formation rate in our 182Mpc/hr
  runs in comparison to different observations and other
  theoretical predictions.
  } 
\label{sSFR}
\end{figure}

Fig. \ref{sSFR} depicts the redshift evolution of the mean specific
SFR for our two 182Mpc/hr runs. As in \citet{Biffi} -- who studied early
proto-galaxies at $z > 9$ -- we compare our simulations with other
theoretical models (i.e. \citealt{Biffi},
\citealt{Dayal}, \citealt{Dave}) and observations
(i.e. \citealt{Noeske}, \citealt{Daddi}, \citealt{Dunne},
\citealt{Pannella}, \citealt{Stark09}, \citealt{Yabe},
\citealt{Michalowski}, \citealt{Schiminovich}, \citealt{Reddy},
\citealt{Bouwens}, \citealt{Gonzalez}, \citealt{Zheng_nature},
\citealt{Stark13} and \citealt{Coe}). 
Irrespectively of the assumed accretion and feedback models, our
simulations are both in better agreement with observations than many
other theoretical models, especially at low redshifts (where the
observational constraints are tighter).
Fig. \ref{sSFR} also demonstrates that our new implementations have no 
effect on the specific SFR. Hence, the changes in the SFR and in the stellar mass are the same.

However, star formation is certainly not only regulated by AGN
feedback. Recent studies (e.g. \citealt{Hopkins_2013},
\citealt{Hirschmann_Naab}, \citealt{Aumer}, \citealt{Kannan}) showed
that stellar feedback also plays an important role, particularly for
low mass galaxies. 
\begin{figure}
  \includegraphics[trim = 0mm 0mm 0mm 5mm, clip, width=0.5\textwidth]{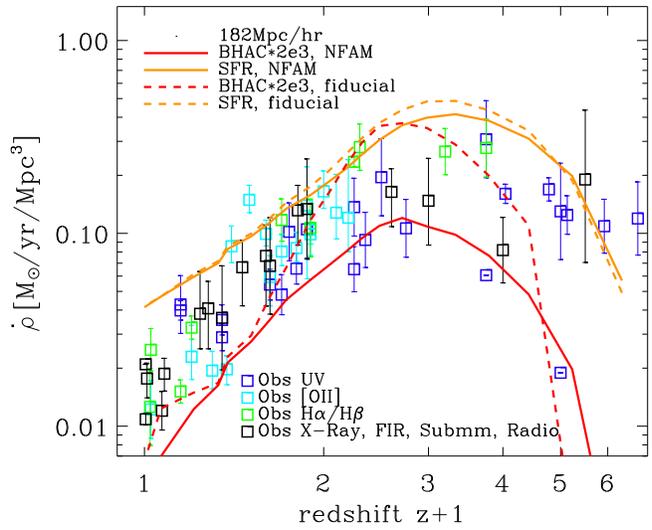}
\caption{History of the star formation (orange lines) and black hole
  accretion rate (red lines) density in both 182Mpc/hr runs (fiducial
  model: dashed lines, NFAM: solid lines) in comparison to
  observations from \citealt{Hopkins_Beacom} (squares).
  } 
\label{SFH}
\end{figure}
Fig. \ref{SFH} provides further evidence that our model is still not
sufficient for reproducing galaxies with realistic SFRs.
It illustrates the history of the star formation and the black hole
accretion rate densities as shown by \citet{Hirschmann} for our two
182Mpc/hr runs compared to observations of the SFR density (squares) by 
\citet{Hopkins_Beacom}.
In comparison to the fiducial model, the star formation rate density
in the NFAM model is slightly lower above $z\approx 1.5$, although it
is still too high in comparison to the observations except for very
high redshifts, which are, however, affected by resolution. 

As expected due to the lower black hole masses in the NFAM model, the
black hole accretion rate density is significantly lower at $z<4.5$
than in the fiducial model. For higher redshifts, it is larger than in
the fiducial model, which leads to a much shallower increase up to the
maximum. 
Fig. \ref{SFH} demonstrates that in the NFAM simulation the SFR and
the black hole accretion rate evolve very similar with redshift. The
reason is that both depend on the amount of cold gas. With our new
accretion model the analogy between SFR and black hole accretion is
even stronger, because the accretion factor for hot gas is smaller
than for cold gas. 
Thus, in the NFAM simulation, hot gas results not only in less star
formation, but also in smaller black hole accretion rates.
This shows that the gas temperature plays a key role in
both galaxy formation and black hole growth. A similar accordance
between the history of the star formation and black hole accretion
rate density was also found by \citet{Zheng}, who adopted the
luminosity functions from \citet{Hopkins} to estimate the black hole
accretion rate densities.

\section{Discussion}
\label{discussion}
\subsection{The effect of the feedback model onto the luminosity functions}
\label{LFs}
As already mentioned before, the choice of the slope $\beta$ of the
feedback model should not have a significant influence on the
resulting galaxy and black hole properties in the simulations since
$\epsilon_{\mathrm{r}}$ is much smaller than $\epsilon_{\mathrm{o}}$. 
However, it has an influence on the AGN luminosity functions, which
are calculated during post-processing using the accretion rates
calculated by the simulation and the radiative efficiencies, which can
be varied.

\begin{figure}
  \includegraphics[trim = 5mm 3mm 10mm 10mm, clip, width=0.5\textwidth]{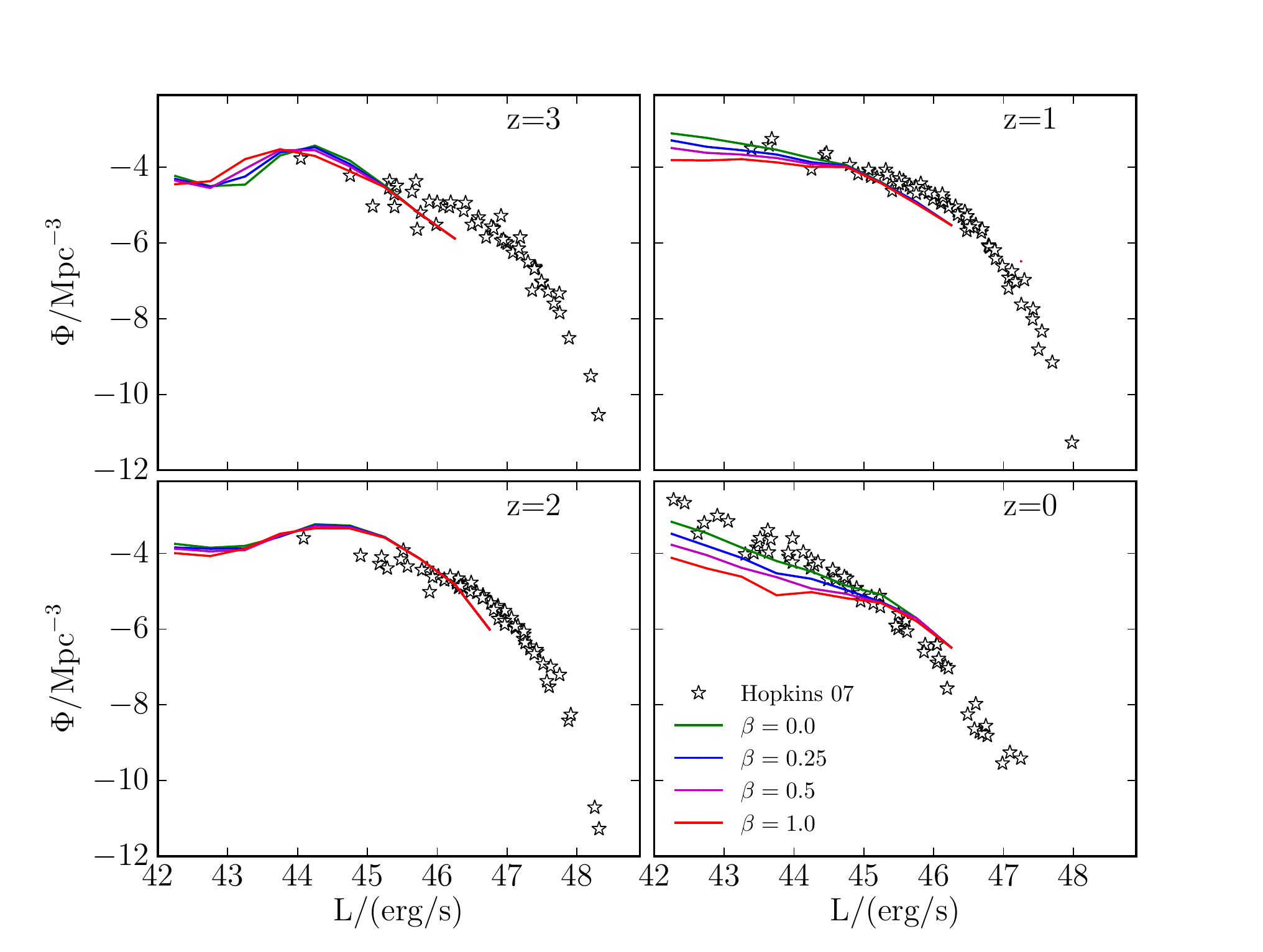}
  \caption{AGN luminosity function of our 182Mpc/hr NFAM run at
    different redshifts for different values for the slope $\beta$ in
    comparison to the observational compilation by \citet{Hopkins}.
    } 
\label{LF_slope}
\end{figure}

In that way we can test the effect of the parameter $\beta$ on the 
AGN luminosity function.
We calculate the bolometric AGN luminosities
of the NFAM simulation for different values of $\beta$ using equation
(\ref{radiation_power}) and (\ref{epsilon_r_final}). 
Fig. \ref{LF_slope} shows the resulting luminosity functions in
comparison to the observational compilation of \citet{Hopkins}. 
For a comparison of moderately luminous AGN, particularly at high
redshifts, one has to keep in mind that simulations are affected by
resolution (see discussion of \citealt{Hirschmann}). 
In addition, dust obscuration effects in observational data typically
result in an underestimation of their number density
(e.g. \citealt{Hasinger08}, \citealt{Merloni14}) which complicates a
comparison between simulations and observations. 
Even if luminosity-dependent obscuration effects on a torus level are
already considered in \citet{Hopkins}, an additional
redshift-dependence (of X-ray luminosities, as suggested by
e.g. \citealt{Hasinger08} and \citealt{Merloni14}) may change the low
luminous end at high redshifts. 

Fig. \ref{LF_slope} shows that the effect of the choice of $\beta$ on
the AGN luminosity functions is not significant, especially at high
redshifts, because $\beta$ changes only the efficiencies in the
radio-mode and not in the quasar-mode. For lower redshifts, when more
black holes accrete with low Eddington ratios, it has an influence on
the amount of 
AGN with luminosities smaller than $10^{45}$erg/s in the sense that
with decreasing $\beta$ the radiative efficiency and thus the amount
of moderately luminous AGN is increasing and thus the result is in
better agreement with the observational constraints. 
However, due to the fact that observations constrain very low values
of $\epsilon_{\mathrm{r}}$ we suspect that the accretion rates in the
quasar-mode are slightly underestimated in our simulations. 

As shown in Fig. \ref{feedback_model_plot}, the actual value of
$\epsilon_\mathrm{r}$ is entirely unconstrained in the radio regime. 
It might depend on many properties like the morphology of the host
galaxy or the merger history of an individual black hole. 
For that reason, calculating a more realistic value of
$\epsilon_\mathrm{r}$ is beyond the current feasibility. 
\begin{figure}
  \includegraphics[trim = 5mm 3mm 10mm 10mm, clip, width=0.5\textwidth]{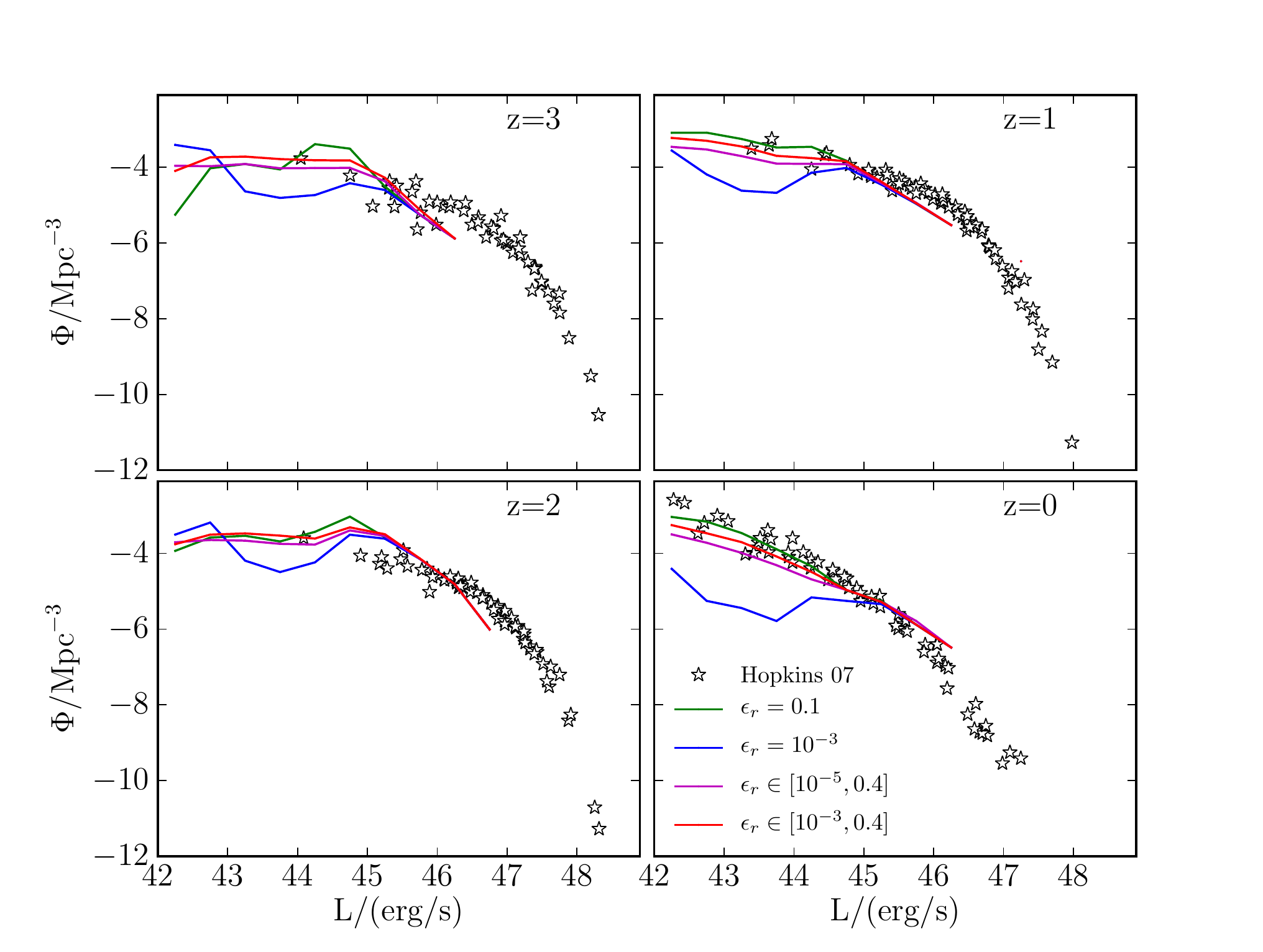}
  \caption{AGN luminosity function of our 182Mpc/hr NFAM run at
    different redshifts for different values of $\epsilon_\mathrm{r}$
    in the radio regime in comparison to the observational compilation
    by \citet{Hopkins}. The green and blue curves show the
    result for two constant values of $\epsilon_\mathrm{r}$. For the
    purple and red curve we took random values in two different
    intervals.
    } 
\label{LF_models}
\end{figure}

Nevertheless, according to the observations by \citet{Russell}, one
should consider different models to estimate  $\epsilon_\mathrm{r}$ in
the radio-mode. Fig. \ref{LF_models} shows the AGN luminosity
functions in comparison to observational compilation by \citet{Hopkins}
for four models adopting different values for $\epsilon_\mathrm{r}$ in
the radio regime: 
\begin{enumerate}
\item{The commonly used value  $\epsilon_\mathrm{r}=0.1$ (green lines)
    seems to match the observations reasonably well, although such a
    value is unlikely according to the results from \citet{Russell} and
    \citet{Mezcua}.} 
\item{$\epsilon_\mathrm{r}=10^{-3}$ is the mean value of the data
    points from \citet{Russell}. Because we change only values in the
    radio regime, the high luminosity end is not affected. At lower
    luminosities, the AGN number densities are significantly
    underestimated as AGN become way too faint\footnote{The amount of
      black holes does not change, because we use the same simulations
      for all different feedback models. Consequently, lower number
      densities of AGN with $L > 10^{42}$erg/s are equivalent to higher
      number densities for fainter AGN.} (blue lines).} 
\item{We choose random values in log space in the range
    $10^{-5}<\epsilon_\mathrm{r}<0.4$. This is approximately the range
    of the data points from \citet{Russell} with a maximum value equal
    to the theoretical maximum efficiency of a rotating black hole
    (since we assumed $\eta=0.1$). It leads to a reasonably good match
    (magenta line) with the observational constraints, even if the low
    luminous end is slightly lower than when adopting the commonly
    used value (green lines). Since we may speculate that the curve
    will probably be shifted upwards when choosing a higher resolution
    \citep{Hirschmann}, the concordance with the observations might be
    even better.}  
\item{Now we exclude very low values for $\epsilon_\mathrm{r}$ and
    hence choose random values in the range
    $10^{-3}<\epsilon_\mathrm{r}<0.4$. This leads to a slightly, but
    not significantly larger number density of moderately luminous AGN
    (red lines) and hence to a better agreement with observations.}
\end{enumerate}
In comparison to the AGN luminosity functions from the Illustris
simulation \citep{Sijacki_2014}, we have less luminous AGN for
redshifts below $z=1$, 
although our cosmological box is larger.
Nevertheless, to investigate the high-mass end in more detail larger
cosmological boxes are needed. 
\citet{Hirschmann} already presented luminosity functions of a larger
box from the set of Magneticum Pathfinder Simulations,  
which are in good agreement with the observations from \citet{Hopkins}.
Furthermore, our simulation matches better with the observed amount of
AGN with luminosities below $L \approx 10^{45}$erg/s than in
\citet{Sijacki_2014}. 
This confirms the conclusion from \citet{Sijacki_2014} that the
radiative efficiency is not constant and might actually be very low in
the radio regime. 

This analysis shows that the efficiency of the radiative
component in the radio regime is indeed not yet understood because the
theoretical lower limit is not captured by observations. 
Interestingly, choosing random values for the radiative efficiency in
the range of the observed values leads to a good agreement with
observed AGN luminosity functions. 
This may indicate that in the radio regime the radiative efficiency
depends neither on the mass of the black hole, nor on its accretion
rate. It also implies that -- as we are matching the observed
luminosity function by randomly choosing the radiative efficiency
within the observed values -- the distribution of the accretion rates
as predicted by the simulations are similar to the observed ones.  
We conclude, that it is theoretically not fully understood how
efficient AGN radiate and we suspect that the morphology of the
galaxy, but also turbulence or even magnetic fields might play an important
role. Since jets dominate in the radio-mode, they can also prevent
efficient accretion.
The similar morphologies of the two radiation dominated sources from
\cite{Mezcua}, i.e. NGC 1097 and NGC 1386, give a first evidence for
these speculations, 
because they both have a ring of star forming regions and a bar on
large scales, but no bar on small scales. 
However, a better understanding of black hole accretion and AGN
feedback processes is a 
great challenge for the future, because more accurate observations are 
needed
to learn in which cases ADAF/Bondi models are a good estimate and in
which cases we have to include additional physical processes. 

\subsection{The unconstrained total efficiency in the radio regime}
\label{The_unknown}
Besides the radiative efficiency, the total efficiency $\eta$ in the
radio regime is also unconstrained. Throughout this study, we always assumed
$\eta=0.1$ to calculate $\epsilon_\mathrm{r}$ and
$\epsilon_\mathrm{o}$, making, thus, our conclusions for the radio
regime rather uncertain.
The reason for this assumption are missing or unconstrained
estimations of $\dot{M_\bullet}$. 
According to equation (\ref{Mdot_Edd_Toy}), $\eta$ is given by
\begin{equation}
\eta = \frac{L_{\mathrm{Edd}}}{\dot{M}_{\mathrm{Edd}} c^2} =
\frac{L_{\mathrm{bol}}
  \frac{\dot{M_\bullet}}{\dot{M}_{\mathrm{Edd}}}}{\frac{L_{\mathrm{bol}}}{L_{\mathrm{Edd}}}
  \dot{M_\bullet} c^2}. 
\label{eta}
\end{equation}
In observations, however, usually only the AGN luminosity, the jet
power and the black hole mass are measured. Using the black hole mass,
one can calculate $L_{\mathrm{Edd}}$. Equation (\ref{mdot_splitted})
is then used to calculate
$\dot{M_\bullet}/\dot{M}_{\mathrm{Edd}}$. Hence $\dot{M_\bullet}$ is
the parameter which is typically missing. 
Nevertheless, for some of the sources from \citet{Russell} and
\citet{Mezcua}, $\dot{M_\bullet}$ has been estimated. 
We use these estimations to calculate the corresponding total
efficiencies with equation (\ref{eta}). 
With these values and equations (\ref{epsilon_o_obs}) and
(\ref{epsilon_r_obs}) we then compute $\epsilon_{\mathrm{o}}$ and
$\epsilon_{\mathrm{r}}$. 

\begin{figure}
  \includegraphics[trim = 4mm 0mm 22mm 10mm, clip, width=0.5\textwidth]{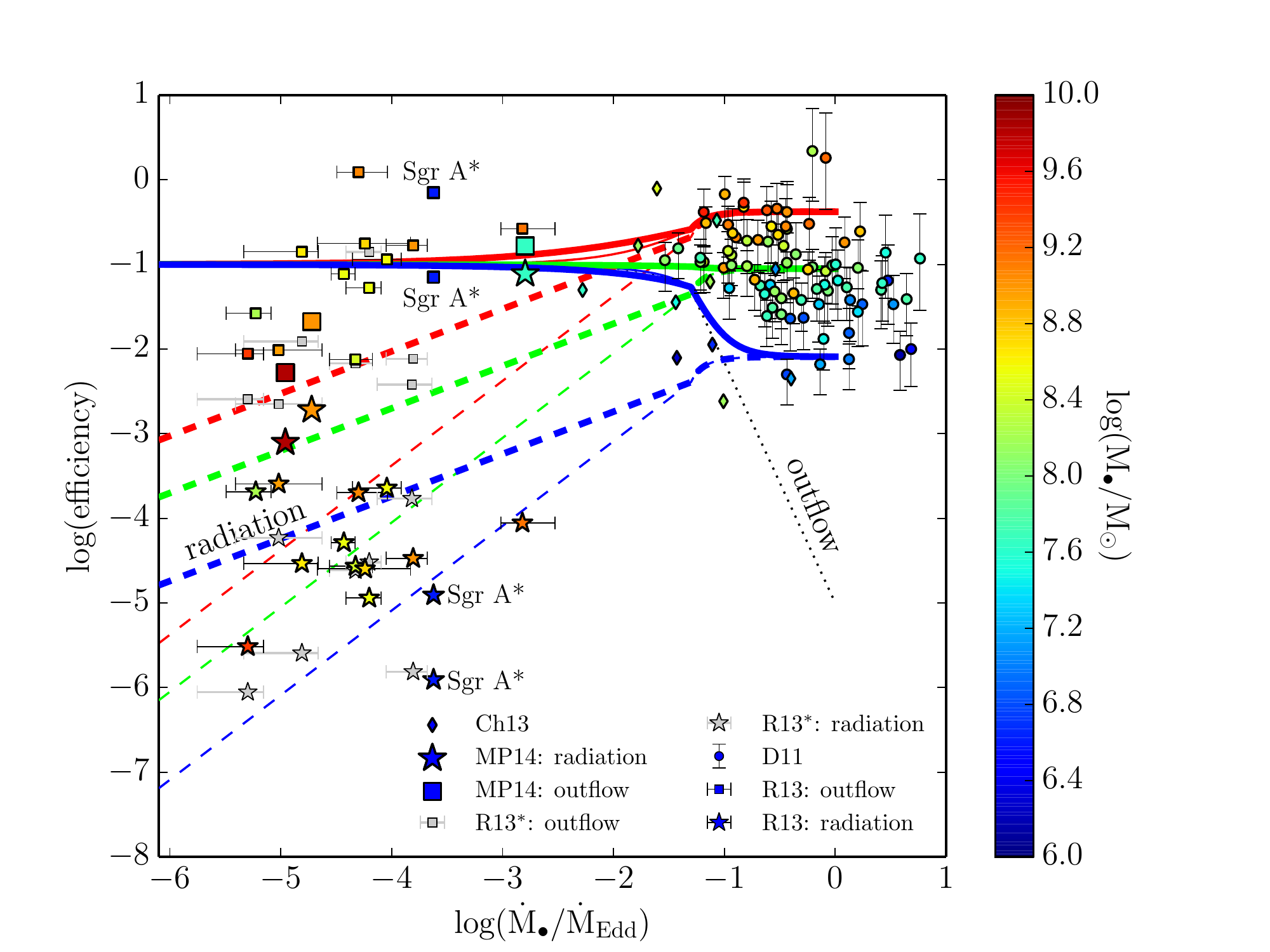}
  \caption{Same as in Fig. \ref{feedback_model_plot}, but with
    efficiencies calculated using values for $\dot{M_\bullet}$ from
    \citealt{Russell} (R13) and from other authors, i.e. \citealt{Evans},
    \citealt{Allen} and \citealt{Li} (R13$^*$, MP14). The three data points
    from \citet{Mezcua}, for which we know estimations of $\dot{M_\bullet}$
    are from left to right M87, NGC 4594 and CenA.
    We also included values for Sgr A*, which have been calculated using
    different estimations of $\dot{M_\bullet}$, i.e. $\dot{M} \approx
3 \cdot 10^{-6} M_{\odot}/\mathrm{yr}$ from \citealt{Cuadra} (upper symbols) and $\dot{M} \approx
3 \cdot 10^{-5} M_{\odot}/\mathrm{yr}$ from \citealt{Quataert} (lower symbols).} 
\label{feedback_model_plot_real}
\end{figure}

Before we calculate the efficiencies for the selected sources, we want
to focus on the nearest SMBH, namely Sagittarius A* (Sgr A*). 
For the luminosity we adopt $L_{\mathrm{bol}}=2.1 \cdot 10^{36}
\mathrm{erg/s}$ \citep{Narayan98} and for the power of the mechanical
outflow we assume $P_{\mathrm{o}}=1.2 \cdot 10^{41} \mathrm{erg/s}$
\citep{Yusef}. With these values and the mass $M_{\mathrm{SgrA*}} = 4 \cdot 10^6
M_{\odot}$ we calculate the Eddington ratio using equation
(\ref{mdot_splitted}). Although Sgr A* is the nearest SMBH, there are
different estimates for the accretion rate. \citet{Quataert99}
estimated a Bondi accretion rate of $\sim 3 \cdot 10^{-5}
M_{\odot}/\mathrm{yr}$. However, there are other models suggesting the
actual accretion rate might be much lower than the Bondi accretion
rate (e.g. \citealt{Quataert}). \citet{Cuadra} derived $\dot{M} \approx
3 \cdot 10^{-6} M_{\odot}/\mathrm{yr}$ from stellar winds. 
We calculated the efficiencies corresponding to both values using
equation (\ref{epsilon_o_obs}) and (\ref{epsilon_r_obs}). 
They are shown in Fig. \ref{feedback_model_plot_real}.
The upper data points belong to $\dot{M} \approx 3 \cdot 10^{-6}
M_{\odot}/\mathrm{yr}$ and the lower ones to $\dot{M} \approx 3 \cdot
10^{-5} M_{\odot}/\mathrm{yr}$. 
Assuming that the ADAF model really provides a lower limit, this
illustrates that $\dot{M} \approx 3 \cdot 10^{-6}
M_{\odot}/\mathrm{yr}$ is in good agreement with our model for the
radiative efficiency. It also indicates that it is necessary to choose
different lower limits for different black hole masses,
because the dashed green line -- which corresponds to $\eta \approx
0.1$ -- is far above the data point.  
However, the corresponding value for $\epsilon_\mathrm{o}$ is larger
then the commonly used value 0.1.  
This indicates, that the outflow efficiencies might differ
significantly from this value, which is not well constrained.  
For the second estimation of the accretion rate, i.e. $\dot{M} \approx
3 \cdot 10^{-5} M_{\odot}/\mathrm{yr}$, the 
radiative efficiency is clearly below the prediction, although
$\epsilon_\mathrm{o}$ is near 0.1. This implies that Bondi
estimations of the accretion rate indeed tend to be too high. 

Now, we consider the sources from \citet{Russell} and \citet{Mezcua},
for which $\dot{M_\bullet}$ has been estimated using the Bondi
model. \citet{Russell} investigated a subsample of 13 objects for which
they estimated $\dot{M_\bullet}$. The efficiencies corresponding to
these sources are plotted in Fig. \ref{feedback_model_plot_real}
(R13). Other authors also estimated $\dot{M_\bullet}$:
for Centaurus A and NGC 4216 we use the result from \citet{Evans} and
for the Sombrero galaxy (NGC 4594) we take $\dot{M_\bullet}$ from
\citet{Li}. For M87, M84, M89, NGC 4636, NGC 4472, NGC407 and NGC5846
we take values from \citet{Allen}. The efficiencies calculated with
these values and the data from \citet{Russell} are marked with grey
symbols (R13$^*$). Most of these sources are also in the selected sample
from \citet{Russell}. We can, thus, directly compare the results of two
independent measurements. This shows a clear discrepancy between
different estimations of $\dot{M_\bullet}$. Overall, the efficiencies
are larger using the $\dot{M_\bullet}$ from \citet{Russell}. 
In contrast to Fig. \ref{feedback_model_plot}, the lowest values of the
radiative efficiency now tend to increase with increasing Eddington
ratio as predicted by theory. Nevertheless, the observations are in
better agreement with theory using only the 13 objects of the
selected subsample. Furthermore, Fig. \ref{feedback_model_plot_real}
indicates that the value $\epsilon_{\mathrm{o}}=0.1$ is indeed a
reasonable assumption for the mean value of the observed values,
although the observations can be nearly two dex lower. 

However, all these estimations are highly uncertain and very speculative.
On the one hand, all data points are upper limits due to the
approximation of using the Bondi model. 
On the other hand, there are studies showing that accretion rates can
also be much smaller than $\dot{M}_\mathrm{B}$
(i.e. \citealt{Li_Ostriker_Sunyaev}, \citealt{Baganoff},
\citealt{Quataert}).  Moreover, values for $L_{\mathrm{bol}}$ might be
underestimated when the jet is emitting in the plane of the sky. 
In that case, the measured flux is smaller than if the jet were located
close to the line of sight. This would lead to higher radiative
efficiencies and to an even better agreement with our model.
Furthermore, uncertainties in the determination of black hole masses make it almost
impossible to investigate whether the lower limit for the radiative
efficiency splits up for different black hole masses as seen in the
quasar regime (\citealt{Davis},\citealt{Chelouche}). 

Nevertheless, the data shown in Fig. \ref{feedback_model_plot_real} is
one of the best constrained samples. The comparison between
Fig. \ref{feedback_model_plot_real} and Fig. \ref{feedback_model_plot}
shows that we need more accurate measurements to learn more about the
feedback of radio jets and the corresponding efficiencies. Due to the
fact that knowing the efficiencies is (at least with the currently
available computational power) essential for performing large-scale
cosmological simulations, it is worth and necessary spending more
effort on observational estimates of black hole accretion rates.  

\subsection{Comparison with other simulations}
During the last couple of years, several other groups have
  also been working on large cosmological simulations including
  baryons and black holes.
  As our simulations, some of these simulations, for example
  the MassiveBlack-II simulation
  (\citealt{Khandai}), earlier simulations from
  \citet{DiMatteo08} and the new EAGLE simulation
  (e.g. \citealt{Schaye}), are based on the SPH code GADGET-3,
  but differ in their physical sub-resolution
  models, including the model for black hole growth. In contrast, the recent
  Illustris simulation (e.g. \citealt{Vogelsberger14}, \citealt{Genel}) has been performed with a different
  hydrodynamic scheme, the moving mesh code AREPO
  \citep{Springel_Arepo}, and also slightly different sub-resolution
  models. A comparison between these models can help to understand
  which effects the different sub-resolution models for black hole growth
  and AGN feedback may have on basic galaxy and black hole
  properties. 

\begin{figure}
  \includegraphics[trim = 0mm 0mm 0mm 5mm, clip, width=0.5\textwidth]{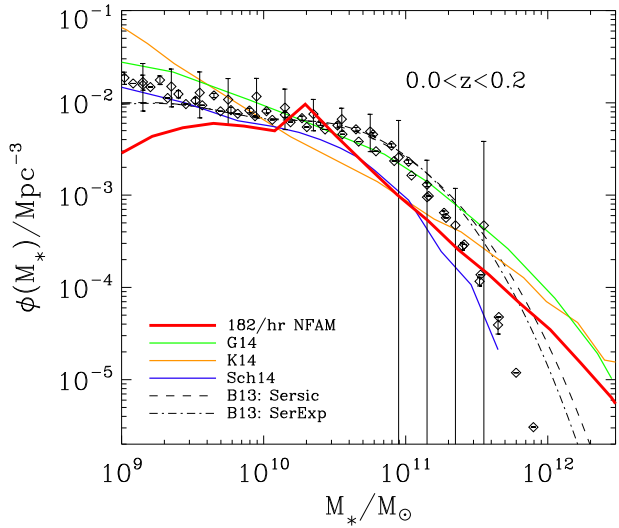}
  \caption{Comparison of the SMF in the NFAM model (red line)
      at $z=0$ with the Illustris simulation (\citealt{Genel}/G14,
      green line), the MassiveBlack-II simulation
      (\citealt{Khandai}/K14, orange line) and the EAGLE simulation
      (\citealt{Schaye}/Sch14, blue line).
      The observations shown are the same as in Fig. \ref{SMF}.
  }
\label{SMF_comparison}
\end{figure}

Fig. \ref{SMF_comparison} shows the stellar mass function in
  the NFAM model below $z=0.2$ in comparison to other simulations.
As for the black hole mass function, the number density of massive
galaxies in the Illustris simulation (\citealt{Genel}, green lines) is
by half an order of magnitude larger than the one in the Magneticum
simulation.
 For stellar masses below $4 \cdot 10^{11} M_{\odot}$ the galaxy
 number densities in the Illustris simulation are in reasonably good
 agreement with the observations, while our simulations produce
 slightly too few low mass galaxies.
Since the difference between the SMFs of the fiducial model and the NFAM model are very small at $z=0$ we suggest that other physical processes (e.g. stellar feedback or cooling) or the lower resolution might be the reason for the lower stellar masses. 
The prediction from the MassiveBlack-II simulation (\citealt{Khandai}, orange line) has no pronounced
exponential cut-off with the consequence that they over-estimate the
low and the high mass end, but slightly under-estimate the number
density of galaxies around the exponential cut-off.
In contrast, the stellar mass function obtained by the EAGLE simulation
(\citet{Schaye}, blue line), where the feedback is especially
calibrated to match the stellar mass functions, is in good agreement
with observations for the entire stellar mass range.

\begin{figure}
  \includegraphics[trim = 5mm 10mm 20mm 120mm, clip, width=0.5\textwidth]{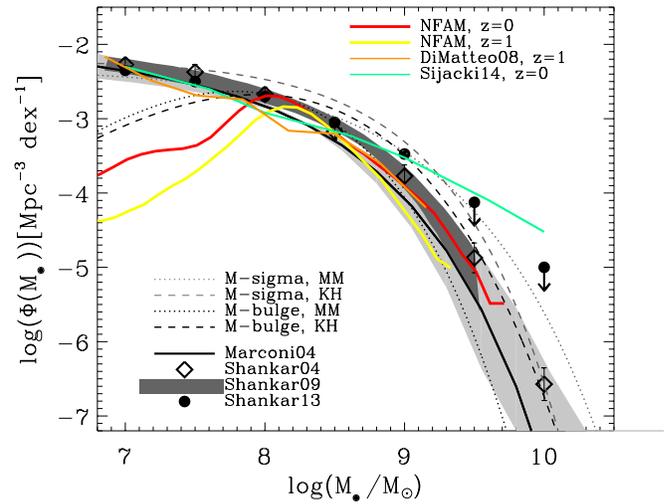}
  \caption{Comparison of the black hole mass function in the
      182Mpc/hr NFAM run with that in the Illustris simulation
      \citep{Sijacki_2014} at $z=0$ and with that in the
      MassiveBlack-II simulation \citep{DiMatteo08} at $z=1$. 
      The observations shown are the same as in Fig. \ref{BMF}.
  }
\label{BHMF_comparison}
\end{figure}

Compared to our results -- the black holes in  the Illustris
  simulation are much more massive than in the Magneticum simulation
  (as shown in Fig. \ref{BHMF_comparison}). This discrepancy might
  have several reasons, for example the different implementations of
  radiative AGN feedback. Furthermore, given that there may still be
  resolution dependent details of the black hole feedback model
  (e.g. the estimation of the Bondi accretion rate or the distribution of the
  feedback) the higher resolution of the Illustris simulation could
  contribute to these differences. In addition, there could be
  differences due to the different numerical techniques,
  namely SPH and moving mesh, especially in the way the feedback gets 
  transported away from the centre of the galaxies.
  In addition, a more efficient gas cooling in AREPO \citep{Nelson_2013}
  might lead to higher black hole accretion rates.
Furthermore, the underlying physics referring to the energy transport
might influence how much gas is driven outward and which fraction of
this gas is recycled as for instance discussed by \citet{Nelson}.

Due to the large uncertainties in different observational estimates it
is not clear which simulation matches the observations of the local
Universe best. At $z=1$ we also compare the black hole mass function
of our NFAM model to the predictions of \citet{DiMatteo08}. This
simulation produces slightly more massive black holes than the
Magneticum simulation, which might be due to a more inefficient AGN
feedback of massive black holes in \citet{DiMatteo08}.

Obviously, the other simulations shown here capture black
  holes down to smaller black hole masses. Firstly, this is due to
  the higher resolutions.
  Secondly, they use the
  so-called 'pinning' to keep the black holes at the potential minimum
  and therefore in the centre of the galaxies. Hence, they can seed 
  the black holes in less massive galaxies. In our simulations this is
  not possible, because the black holes in less well defined galaxies
  would not be able to stay in the centre of their host galaxy due to
  numerical effects. However, not using the so-called 'pinning' avoids
  other drawbacks of this method as discussed in \cite{Hirschmann}.
  As discussed by
  \citet{Shankar13}, also the low mass end of the black hole mass
  function is relatively uncertain and depends on the black hole
  scaling relations. For example, the low mass end could be significantly
  smaller when excluding galaxies with pseudo-bulges.
  Therefore, it will be quite challenging to compare observed black
  hole mass functions to any simulation at the low mass end. 

\begin{figure}
  \includegraphics[trim = 4mm 0mm 10mm 10mm, clip, width=0.5\textwidth]{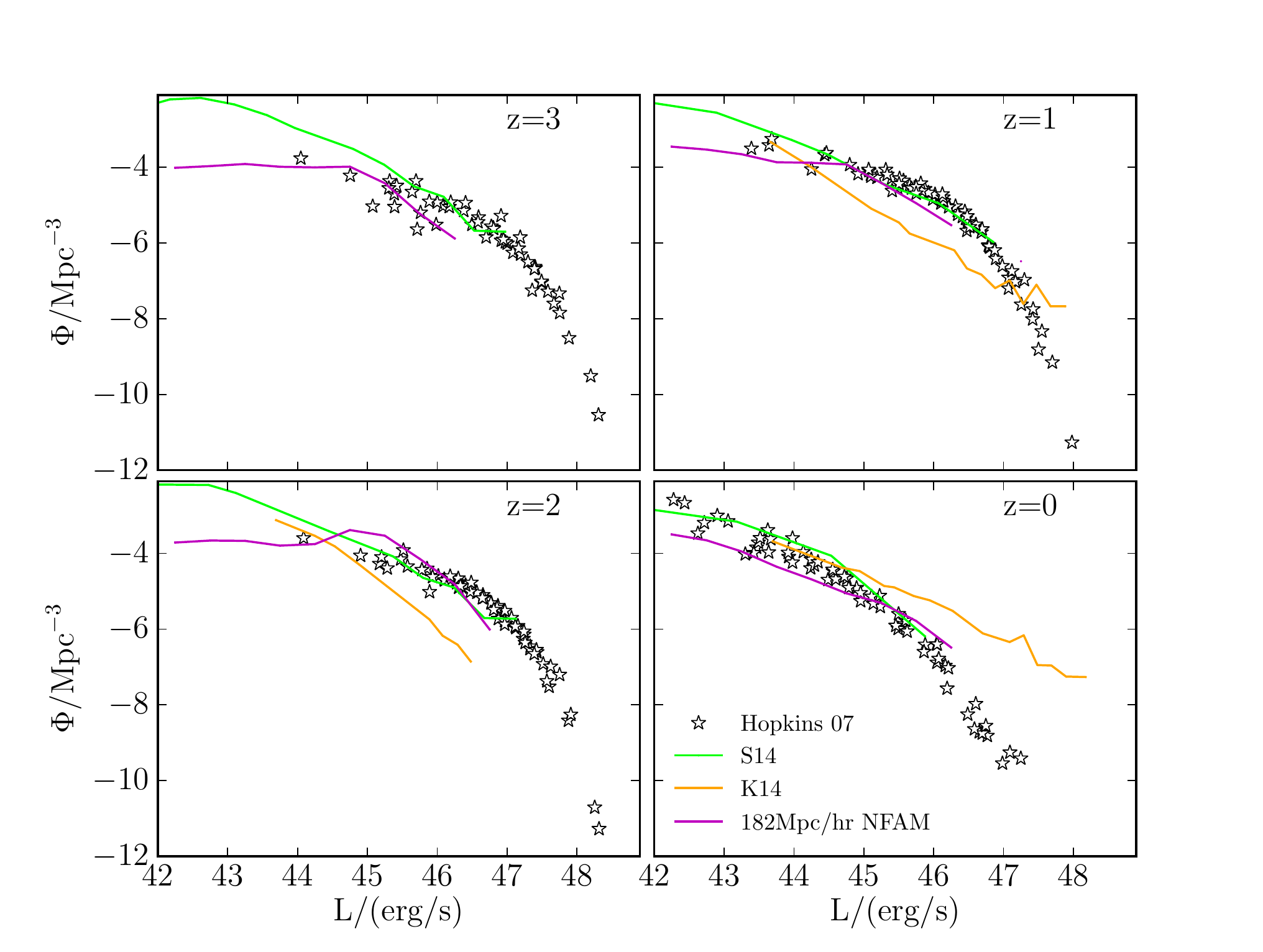}
  \caption{Comparison of the AGN luminosity function in the
      NFAM model (using random radiative efficiencies in the radio
      regime in the range $10^{-5}<\epsilon_{\mathrm{r}}<0.4$) with
      the predictions of the Illustris simulation
      (\citealt{Sijacki_2014}/S14, green solid lines) and of the MassiveBlack-II
      simulation (\citealt{Khandai}/K14, orange solid lines).
  }
\label{LF_comparison}
\end{figure}

  Fig. \ref{LF_comparison} shows a
  comparison of the AGN luminosity function in our NFAM run (purple
  line) with the predictions from the Illustris simulation
  (\citealt{Sijacki_2014}, green solid line) and from the
  MassiveBlack-II simulation (\citealt{Khandai}, orange solid
  line). The luminosity function of the Illustris simulation matches
  both the  observations and our simulation, whereas the
  MassiveBlack-II simulation widely fails to reproduce the observed 
  shape of the observed luminosity functions of \citet{Hopkins}.
  Since the latter simulation contains the original model from
  \citet{Springel_BHs} with only one mode of AGN feedback, we can
  speculate that this might be one possible reason for the
  discrepancies.
  The Illustris simulation
  uses a so-called 'radiative' efficiency, which is implemented as a
  change in the net cooling rate and is most efficient in the
  quasar-mode \citep{Sijacki_2014}.  This seems to have a similar
  effect as our variable radiative efficiency, which increases for
  large black hole masses in the quasar mode. 

Nevertheless, we want to emphasize that despite of the general
importance for understanding the (physical or numerical) origin of
different simulation predictions, such a comparison must remain
speculative: besides different models for black hole growth and AGN
feedback, many other physical details (e.g. models for star formation,
stellar feedback) or different hydrodynamic schemes may cause more
fundamental changes in basic galaxy properties. Such an investigation
is, however, clearly beyond the scope of this work.

\section{Summary and conclusions}
\label{Summary}
In this paper, we presented an improved implementation of the black
hole model originally introduced by \citet{Springel_BHs}.
We combined theoretical predictions of \citealt{Churazov},
\citealt{Narayan} and  \citet{Gaspari} with observations from
\citealt{Russell}, \citealt{Mezcua}, \citealt{Davis} and
\citealt{Chelouche} in order to  model the underlying sub-grid
processes more realistically.   

The new model includes a combination of mechanical outflow and
radiation, which we both implemented as thermal feedback due to the
inability of resolving sub-kpc scales, where jets provide the
mechanical feedback. Both feedback processes are modelled as a 
function of the actual accretion rate with respect to the Eddington
rate, which leads to a smooth transition between the outflow-dominated
radio-mode and the radiation-dominated quasar-mode. In addition, our
model includes a mass dependent radiative efficiency to account for
the observed spin of the black holes.  

Furthermore, we distinguish between the hot and the cold gas component  
within the environment of the black holes and calculate the accretion
rate for these two components separately. This allows us to model the
Bondi accretion differently for the two phases, where we use two
different boost factors ($\alpha = 10$ for the hot and $\alpha=100$
for the cold gas) according to the results of small-scale simulations of
\citet{Gaspari}.  

Besides that, free parameters of the model (like the various
efficiencies) are now more strictly linked to values inferred from
observations. Compared to the fiducial model, our new implementations 
predict a more realistic population of black holes and their host
galaxies, when compared to fundamental observational constraints, in
several aspects: 
\begin{enumerate}
\item{The slope and normalization of the produced $M_{\bullet}$-$M_*$
    relation are in much better agreement with observations over a larger
range of galaxy masses and redshifts than in the fiducial model.
In particular, these improvements are due to the faster black hole
growth at large redshifts and the lower black hole masses at the
massive end for redshifts below $z\approx 2$.}  
\item{Our new feedback scheme is also able to efficiently suppress the 
    late growth of massive black holes. Hence, the resulting
    present-day black hole mass function provides an excellent match
    to the observed one.
    }
\item{In the NFAM simulations, the equilibrium between gas cooling and
    AGN feedback within the galaxies is reached earlier. Consequently,
    star formation starts to be suppressed at earlier times. This
    leads to a better agreement with observed stellar mass functions
    than before.
    In particular, in the NFAM simulation there are much more quiescent galaxies at high redshifts than in the fiducial simulation,
    in which galaxies become quiescent far too late.
    However, some inconsistencies between observed and simulated SMFs for quiescent and star-forming galaxies remain.
}  
\item{The baryon conversion efficiencies are more consistent with
    observations and abundance matching predictions than before,
    although they are still too high by a factor of 2-3 at very high
    stellar masses.} 
\end{enumerate}

A comparison with other large cosmological simulations
(e.g. Illustris, MassiveBlack-II) illustrates that the original
black hole model from \citet{Springel_BHs} needs to be extended to be
able to reproduce observations.  In particular, we find that
\begin{enumerate}
\item{our NFAM simulation successfully matches the observed 
    $M_{\bullet}$-$M_*$ relation. As our fiducial model, the
    simulations from \citet{Sijacki_2014} and \citet{Khandai} do not
    manage to entirely reproduce the observed slope. This may be due
    to the constant values adopted for their radiative efficiencies.  
    }
\item{
    In contrast to the MassiveBlack-II simulation, both our NFAM simulation and the Illustris simulation
    are able to reproduce the observed luminosity functions.
    We suggest that this might be due to the distinction between quasar-mode and
    radio-mode.
    }
\item{our model predicts a lower high mass end of the black hole mass
    functions than other simulations (i.e. \citealt{DiMatteo08},
    \citealt{Sijacki_2014}), because the new AGN feedback model is
    more efficient in limiting black hole growth at higher
    masses. Although all simulations are compatible with the upper
    limits of the black hole mass function estimated from observations
    by \citet{Shankar13}, our model is in excellent agreement with the
    observational data from \citet{Marconi}, \citet{Shankar04}
    and \citet{Shankar09}.  
    }
\item{We predict lower stellar masses than \citet{Genel} and \citet{Khandai}.
    Since our new implementations do not change the SMFs at $z=0$ significantly,
    we suggest that other physical processes like stellar feedback or cooling
    might be the reason for the differences.
    In addition, we find that improvements in the model for star
    formation and stellar feedback like in \citet{Schaye} might be
    necessary to better reproduce the observed shape of the SMFs. 
    }
\end{enumerate}

Despite of the overall success of the NFAM model, open
questions regarding the actual values of the feedback efficiencies
remain.
In contrast to the quasar-mode, the radiative efficiency in the
radio-mode does not show clear trends in observations, which
generally have large uncertainties, especially due to the difficulties
in accurately determining the accretion rate. At high redshifts, the
quasar luminosity function predicted by the simulations is quite
insensitive to the choice of the radiative efficiency in the
radio-mode. However, the best match between simulated and 
observed quasar luminosity functions -- especially at low redshifts --
is obtained when applying a random radiative efficiency to the
simulated AGN in the radio-mode with no dependency on black hole mass
or actual accretion rate. 

Studying the growth of black holes in more detail (i.e. for
individual objects) provides evidence for a two phase process
controlling the evolution of the accretion onto the black hole and the
associated feedback: 
\begin{enumerate}
\item{As long as black holes have masses below the $M_{\bullet}$-$M_*$
    relation, they grow mainly due to continuous gas accretion. This
    phase is primarily driven by cold gas accretion with an accretion
    rate that increases up to the Eddington limit. In this phase, AGN
    are observed as luminous X-ray sources.
This means that the most luminous AGN are not necessarily driven by
merger events as long as they are below the $M_{\bullet}$-$M_*$
relation.}  
\item{When the $M_{\bullet}$-$M_*$ relation is reached, gas cooling and
    AGN feedback are in equilibrium.  Consequently, hot gas accretion
    begins to dominate. This means that the accretion rate, compared
    to the original implementation, is lowered since we correctly
    reduce the boost factor for the hot phase. In this phase, AGN
    feedback is mostly visible as radio jets. This low accretion phase
    can be disturbed by mergers or other processes driving cold
    gas into the centre of the galaxy. In a forthcoming study of the most
luminous AGN in a simulation with higher resolution we will
investigate in more detail whether those objects are mainly triggered
by major mergers.}  
\end{enumerate}
Regarding the latter point, more detailed studies are needed to better
differentiate the AGN triggering mechanisms (as galaxy major and minor
mergers) and their correlation with the black hole accretion processes
within a cosmological context. The next generation of simulations will
also allow to distinguish between morphological types of galaxies
in more detail and thus, to investigate the connection between AGN
luminosities and the host galaxy morphologies, hopefully shedding more 
light on the main trigger mechanisms for AGN activity in different
redshift and luminosity regimes. Such future simulations will also
help to understand the dependency of the AGN driving mechanisms on the  
large-scale environment.  

In addition, we plan to further improve the current implementations by
taking the angular momentum of the accreted material into account,
which in turn would allow to better model the direction of the
feedback. This would especially have an important effect on the
spatial distribution of the feedback energy in the surroundings of the
AGN. 
Indeed, current black hole accretion and feedback models are purely
empirically motivated and have the major drawback that they do not
capture the underlying small-scale physical processes, which is,
within the framework of large-scale cosmological simulation, currently
not feasible due to limited computational power. 
Nevertheless, despite of the rather crude approximations, the black
hole model, in particular with our new modifications, seems to capture
the essence of how black holes grow and how feedback affects the host
galaxies in reality. 

Future observations will improve our understanding of the different
accretion modes and their relation to the multi phase nature of the
ICM/IGM. In particular, studies of Seyfert galaxies (Mezcua et al. in
prep.) will allow an investigation of the role of warm H2 gas (with
temperatures of $\sim 10^3$K). In combination with X-ray observations,
this will shed more light on the complicated interplay between the
various accretion modes of AGN.

\section*{Acknowledgments}
We thank the referee for a careful
and constructive reading of our paper.
We also thank Alexander Beck, Veronica Biffi, Andreas
Burkert, Massimo Gaspari, Mar Mezcua, David Schlachtberger, Francesco Shankar and
Adelheid Teklu for many fruitful discussions. Particularly, we thank Madhura Killedar for carefully
reading the text and editing.
Furthermore we would like to thank Shane W. Davis and Helen Russell
for providing us with observational data and Veronica Biffi
and Umberto Maio for providing us with the data for Fig. \ref{sSFR}.

We are especially grateful for the support by M. Petkova through the
Computational Center for Particle and Astrophysics (C2PAP). 
Computations have been performed at the 'Leibniz-Rechenzentrum' with
CPU time assigned to the Project 'pr86re'. 

This research was supported by the DFG Cluster of Excellence 'Origin
and structure of the universe' and the SFB-Tansregio TR33 'The Dark
Universe'. 

Michaela Hirschmann acknowledges financial support from the European
Research Council under the European Community’s Seventh Framework
Programme (FP7/2007- 2013)/ERC grant agreement n. 202781 and support
from the European Research Council via an Advanced Grant under grant
agreement no. 321323NEOGAL. 

M. Almudena Prieto thanks the hospitality of the
Universit\"ats-Sternwarte M\"unchen and the Max-Planck-Institut f\"ur
extraterrestrische Physik where she stayed as visitor.

\bibliography{modeling_black_holes}

\bsp

\label{lastpage}

\end{document}